\documentclass{article}
\usepackage[right=2.54cm,left=2.54cm,top=2.54cm,bottom=2.54cm]{geometry}
\usepackage{cite}
\usepackage{amsmath,amssymb,amsfonts,amsthm}
\newtheoremstyle{dotless}{}{}{\itshape}{}{\bfseries}{}{ }{}
\usepackage{xcolor}
\usepackage{xurl}
\usepackage{algorithm, algpseudocode}
\usepackage{MnSymbol}
\usepackage{ieeetrantools} 
\usepackage{booktabs}

\usepackage{enumitem}

\usepackage{pgfplots}
\pgfplotsset{compat=newest}

\theoremstyle{dotless}
\newtheorem{dummy}{}

\newtheorem{theorem}[dummy]{Theorem}
\newtheorem{corollary}[dummy]{Corollary}
\newtheorem{lemma}[dummy]{Lemma}

\newtheorem{definition}[dummy]{Definition}
\newtheorem{remark}[dummy]{Remark}

\newcommand{\norm}[1]{\lVert#1\rVert}

\newcommand{\rank}{\textrm{rank}}
\newcommand{\im}{\textrm{im}}
\newcommand{\T}{\top\!}

\newcommand{\blkdiag}{\text{blkdiag}}

\renewcommand{\ker}{\text{ker}}

\newcommand{\See}{\mathbb{S}}
\newcommand{\Pee}{\mathbb{P}}
\newcommand{\Ree}{\mathbb{R}}
\newcommand{\Cee}{\mathbb{C}}

\newcommand{\Nee}{\mathbb{N}}

\newcommand{\Zee}{\mathbb{Z}}
\newcommand{\Xee}{\Ree^{n_x}}

\newcommand{\Fcal}{\mathcal{F}}
\newcommand{\Gcal}{\mathcal{G}}
\newcommand{\Hcal}{\mathcal{H}}
\newcommand{\Jcal}{\mathcal{J}}

\newcommand{\Ncal}{\mathcal{N}}
\newcommand{\Ocal}{\mathcal{O}}
\newcommand{\Pcal}{\mathcal{P}}

\newcommand{\Rcal}{\mathcal{R}}

\newcommand{\signalspaceDT}{\mathfrak{B}(\Zee)}

\usepackage{eso-pic}
\AddToShipoutPictureBG*{%
	\AtPageUpperLeft{%
		\setlength\unitlength{1in}%
		\hspace*{\dimexpr0.5\paperwidth\relax}
		\makebox(0,-0.5)[c]{\begin{tabular}{c c}
				Johan Kon \textit{et al}, A Direct State-Space Realization of Discrete-Time Linear Parameter-Varying Input-Output Models, \\
				In preparation for journal submission, uploaded \today \\
		\end{tabular}}
}}

\begin{document}
\bstctlcite{IEEEexample:BSTcontrol} 

\title{\LARGE \bf A Direct State-Space Realization of Discrete-Time Linear Parameter-Varying Input-Output Models\footnote{This work was supported
in part by by Holland High Tech — TKI HSTM via the PPS
allowance scheme for public-private partnerships; in part by ASML; and
in part by Philips Engineering Solutions. (Corresponding author: Johan
Kon)

Johan Kon is with the Control Systems Technology Group, Eindhoven
University of Technology, The Netherlands (e-mail: j.j.kon@tue.nl).
Roland Tóth is with the Control Systems Group, Eindhoven University
of Technology. He is also with
the HUN-REN Institute for Computer Science and Control, Hungary.
Jeroen van de Wijdeven is with ASML, Veldhoven, The Netherlands.
Marcel Heertjes is with the Control Systems Technology Group,
Eindhoven University of Technology, The Netherlands. He is also with ASML, Veldhoven, The Netherlands.
Tom Oomen is with the Control Systems Technology Group,
Eindhoven University of Technology, The Netherlands. He is also with the Delft Center for Systems and
Control, Delft University of Technology, The Netherlands.}
}

\author{Johan Kon, Roland Tóth, Jeroen van de Wijdeven, Marcel Heertjes, Tom Oomen}

\date{\today}

\maketitle

\section*{Abstract}
A minimal state-space (SS) realization of an identified linear parameter-varying (LPV) input-output (IO) model usually introduces dynamic and nonlinear dependency of the state-space coefficient functions, complicating stability analysis and controller synthesis. 
The aim of this paper is to introduce and analyze a direct SS realization of this IO model that avoids this nonlinear and dynamic dependency, at the cost of introducing a nonminimal state.
It is shown that this direct SS realization 1) is reachable under a coprimeness condition on the coefficient functions of the IO model and a well-posedness condition on the model order, and 2) is never observable but that the unobservable directions converge to zero in a finite amount of steps, i.e., that the realization is reconstructible.
The derived results are illustrated through numerical examples in both the LPV and LTI case.

\section{Introduction}
\label{sec:introduction}
Discrete-time (DT) linear parameter-varying (LPV) input-output (IO) models \cite{Shamma2012, Toth2012} are a powerful model class in identification for capturing time and parameter-related system variations, e.g., due to varying operating conditions. In LPV models, the signal relations are linear, but the coefficients describing these relations are functions of a time-varying scheduling signal $p$ that is assumed to be measurable online. Specifically, an DT-LPV-IO model is represented by a higher-order linear difference equation in terms of previous inputs and outputs, and the coefficients in this difference equation are a function of $p$. The resulting parameter-varying behavior can also embed certain nonlinear characteristics under the correct choice of $p$ \cite{Leith2000a}. 
Data-driven system identification methods using LPV-IO models has been thoroughly developed in the last decades \cite{Bachnas2014,Bamieh2000,Bamieh2002, Previdi2003, Butcher2008, Laurain2010, Mercere2011a, Zhao2012a}.

A state-space (SS) realization of these LPV-IO models enables analysis and controller design using Lyapunov/dissipativity theory\cite{Shamma2012, LMIsLectureNotes,Caverly2019}. For example, such a SS realization can be used to compute the maximum energy amplification ($\ell_2$-gain) of the model using the LPV bounded-real lemma \cite{Apkarian1995}. Such a state-space realization can be either minimal or nonminimal, in which a realization is minimal if there exists no realization with a smaller state dimension \cite{Silverman1969, Gohberg1992}. In a minimal SS realization, all states are relevant for the IO behavior of the SS realization, i.e., all directions in the state space are observable and reachable for some scheduling signal $p$ \cite{Gohberg1992}.
In a nonminimal realization, there are directions in the state space that are irrelevant for the IO behavior of the SS realization, which can complicate analysis and control if these directions are marginally stable/unstable. Consequently, minimal realizations are often preferred over nonminimal realizations. In the linear time-invariant (LTI) setting, minimal realizations can straightforwardly be obtained from an identified transfer function \cite{DeSchutter2000}.

In the LPV case, obtaining a minimal state-space realization of an LPV-IO models alters the way in which the coefficient functions of the realization depends on the scheduling, complicating analysis and design based on this minimal realization \cite{Wollnack2015}. Furthermore, obtaining a minimal realization is complex from an algorithmic and computational point of view. Specifically, minimal realizations can be computed by using canonical state constructions \cite{Guidorzi2003, Weiss2005, Toth2007} or by using the behavioral approach \cite{Toth2010}. For both approaches, the resulting state-space realization generally has coefficient functions that are nonlinear functions of the coefficient functions of the original IO model. Furthermore, the SS coefficient functions generally depend not only on the current value of the scheduling signal $p_k$, but also on its shifted values, e.g., $p_{k+1}$, referred to as dynamic dependence. This nonlinear and dynamic dependence severely complicates the analysis and controller design based on a minimal realization \cite{Wollnack2015}. Only in special cases can these effects be avoided \cite{Toth2012b}. 

In contrast, a SS realization that is potentially nonminimal can directly be constructed from an LPV-IO model without altering the scheduling dependency \cite{Toth2013, Goodwin1984,Pearson2004, DePersis2020}. 
This construction uses the delayed outputs and inputs as a state basis and the resulting realization can directly be specified in terms of the coefficient functions of the LPV-IO model, avoiding the nonlinear and dynamic dependency introduced in minimal realizations.
However, this realization is not guaranteed to be minimal. 

Even though there exist methods to convert LPV-IO models to an LPV-SS realization, at present IO models obtained through identification cannot straightforwardly be used for stability analysis and controller design. The goal of this paper is to illustrate that the direct nonminimal realization can be used for this purpose, and that the nonminimality of this realization is usually not a problem for analysis and control.

The main contribution of this paper is an analysis of the direct nonminimal state-space realization in terms of its reachability and observability properties. This is achieved through the following subcontributions.
\begin{itemize}[topsep=0pt,noitemsep]
    \item[C1] A sufficient condition is derived under which the direct realization is reachable. Namely, that it is reachable for constant scheduling trajectories under a coprimeness condition on the IO coefficient functions and a well-posedness condition on the representation (Section \ref{sec:reachability}).
    \item[C2] It is illustrated that the direct realization is always unobservable, i.e., that there exist states that cannot be observed in the output. Next, it is shown that these unobservable states always decay to zero in finite amount of steps, i.e., the realization is reconstructible, as no unobservable unstable states are introduced (Section \ref{sec:observability}).
\end{itemize}
The developed theory is exemplified through examples. 

It is highlighted that these results are also important for data-driven control in both the LPV as well as the LTI setting, especially for behavioral type of approaches relying state realizations consisting of shifted inputs and outputs \cite{Coulson2019, Dai2023, DePersis2020, Park2009, VanWaarde2022, Verlioek2021}. Specifically, the studied direct nonminimal state-space realization is often used in these methods to characterize the behavior of the system through data, and subsequently analyze stability/dissipativity properties and design controllers directly from data. The reachability and observability properties of this state-space realization play an important role in this field as these are intimately coupled to rank conditions necessary for controller synthesis. As such, a clear understanding of the source of nonminimality is needed to avoid possible misinterpretations, as is shown in the examples.

\subsubsection*{Notation} $z_k \in \Ree^{n_z}$ denotes the value of the signal $z: \Zee \mapsto \Ree^{n_z}$ at time index $k\in\Zee$. $\signalspaceDT$ denotes the set of all signals that are finite at all times, i.e., $\signalspaceDT = \{z \ | \ \norm{z_k} < \infty \ \forall  k \in\Zee \}$. Given a matrix $A\in\Ree^{m\times n}$, $A^\dag$ denotes its pseudoinverse, $\ker\ A$ denotes its kernel (or nullspace), and $(\ker\ A)^\perp$ the orthogonal complement of $\ker\ A$. Given those matrices $A,B$, the block diagonal combination $\begin{bmatrix} A & 0 \\ 0 & B
\end{bmatrix}$ is denoted by $\blkdiag(A,B)$.

\section{Problem Formulation}
\label{sec:problem_formulation}
Consider the multi-input multi-output DT-LPV-IO model $G$ mapping input signal $u: \Zee \mapsto \mathbb{R}^{n_u}$ to output signal $y: \Zee \mapsto \Ree^{n_y}$ according to the difference equation
\begin{equation}
    G : \ y_k = -\sum_{i=1}^{n_a} A_i(p_k) y_{k-i} + \sum_{i=0}^{n_b-1} B_i(p_k) u_{k-i},\label{eq:LPV_IO_DT}
\end{equation}
with coefficient functions $A_i: \Pee \mapsto \Ree^{n_y \times n_y}$, $B_i: \Pee \mapsto \Ree^{n_y\times n_u}$ describing the dependency of \eqref{eq:LPV_IO_DT} on the scheduling signal $p:\Zee \mapsto \Pee \subseteq \Ree^{n_p}$, and $n_a \geq 0, n_b \geq 1$ the order of the IO representation, with $n_a=0$ representing a model in which $y_k$ does not depend on $y_{k-i}$.

For stability/dissipativity analysis and controller/observer synthesis using a model in the form of \eqref{eq:LPV_IO_DT}, it is desired to obtain a state-space realization of \eqref{eq:LPV_IO_DT}. A nonminimal SS realization of \eqref{eq:LPV_IO_DT} can directly be constructed by setting the state equal to the delayed inputs and outputs. Specifically in case $n_a > 0, n_b > 1$, the state is defined as
\begin{subequations}
 \label{eq:maximum_ss_state_def}
 \begin{align}
    x_k &= \begin{bmatrix} \bar{y}_k^\T & \bar{u}_k^\T \end{bmatrix}^\T \in \mathbb{R}^{n_x}, \\
    \bar{y}_k &= \begin{bmatrix} y_{k-1}^\T & \hdots & y_{k-n_a}^\T \end{bmatrix}^\T \in \mathbb{R}^{n_y n_a}, \\
    \bar{u}_k &= \begin{bmatrix} u_{k-1}^\T & \hdots & u_{k-n_b+1}^\T \end{bmatrix}^\T \in \mathbb{R}^{n_u (n_b-1)}.
\end{align}
\end{subequations}
with $n_x = n_y n_a + n_u (n_b -1)$. Given this state vector, the evolution of \eqref{eq:LPV_IO_DT} can be written as
\begin{subequations}
\label{eq:maximum_state_space}
\begin{equation}
\label{eq:maximum_state_space_state}
\resizebox{\textwidth}{!}{$
    x_{k+1}
    \!=\!
    \left[
    \begin{array}{ccccc|ccccc}
        \!-A_1(p_k) \! & \! -A_2(p_k) \! & \! \hdots \! & \! -A_{n_a-1}(p_k)  \! &  \! -A_{n_a}(p_k) & B_1(p_k) \! & \! B_2(p_k) \! & \! \hdots \! & \! B_{n_b-2}(p_k) \! & \! B_{n_b-1}(p_k) \! \\
        I &  & & 0 & 0 & 0 & 0 & \hdots & 0 & 0\\
         & I &  &  & 0 & 0 & 0 & \hdots & 0 & 0\\
        & & \ddots  &  & \vdots & \vdots & \vdots & & \vdots & \vdots \\
        0 &  & & I & 0 & 0 & 0 & \hdots & 0 & 0\\
        \hline
        0 & 0 & \hdots & 0 & 0 & 0 & 0 & \hdots & 0 & 0\\
        0 & 0 & \hdots & 0 & 0 & I &  &  & 0 & 0\\
        0 & 0 & \hdots & 0 & 0 &  & I &  &  & 0\\
        \vdots & \vdots & & \vdots & \vdots & & & \ddots & & \vdots \\
        0 & 0 & \hdots & 0 & 0 & 0 & & & I & 0
    \end{array}
    \right] \!
    x_k
    \! + \!
    \left[
    \begin{array}{c}
        \! B_0(p_k) \! \\
        0 \\
        0 \\
        \vdots \\
        0 \\
        \hline
        I \\
        0 \\
        0 \\
        \vdots \\
        0
    \end{array}
    \right] \!
    u_k,
    $}
\end{equation}
\begin{equation} 
\label{eq:maximum_state_space_output}
\resizebox{\textwidth}{!}{$
    y_k\! =\! \left[ \begin{array}{ccccc|ccccc} 
        \! -A_1(p_k) \! & \! -A_2(p_k) \! & \! \hdots \! & \! -A_{n_a-1}(p_k) \! & \! -A_{n_a}(p_k) \! & \! B_1(p_k) \! & \! B_2(p_k) \! & \! \hdots \! & \! B_{n_b-2}(p_k) \! & \! B_{n_b-1}(p_k) \!
    \end{array}\right] \! x_k
    \! + \! B_0(p_k) u_k,
    $}
\end{equation}
\end{subequations}
thus directly obtaining an SS realization of \eqref{eq:LPV_IO_DT}. 
State-space realization \eqref{eq:maximum_state_space} can be compactly written as
\begin{subequations}
    \label{eq:maximum_ss_ABCD}
    \begin{align}\label{eq:maximum_ss_FcalGcal}
        \left[\begin{array}{c}
            \bar{y}_{k+1} \\
            \bar{u}_{k+1}
        \end{array}\right]
        &= 
        \left[\begin{array}{c|c}
            F_a - G_a A(p_k) & G_a B(p_k) \\
            \hline \hspace{-5pt} 
            0 & F_b
        \end{array}\right] 
        \left[\begin{array}{c}
            \bar{y}_{k} \\
            \bar{u}_{k}
        \end{array}\right]
        + 
        \left[\begin{array}{c}
        G_a B_0 (p_k) \\
        \hline
        G_b
        \end{array}\right]
        u_k
        = \Fcal(p_k) x_k + \Gcal(p_k) u_k, \\
        y_k &= \left[\begin{array}{c|c}
             -A(p_k) & B(p_k)
        \end{array}\right]
        x_k + B_0(p_k) u_k = \Hcal(p_k) x_k + \Jcal(p_k) u_k,
        \end{align}
\end{subequations}
with
\begin{subequations}
\begin{align}
    A &= \begin{bmatrix} A_1 & \hdots & A_{n_a-1} & A_{n_a} \end{bmatrix}: \Pee \mapsto \mathbb{R}^{n_y \times n_y n_a}, \label{eq:maximum_ss_A} \\
    B &= \begin{bmatrix} B_1 &  \hdots & B_{n_b-2} & B_{n_b-1} \end{bmatrix}: \Pee \in \mathbb{R}^{n_y \times n_u (n_b-1)}, \label{eq:maximum_ss_B}
\end{align}
\end{subequations}
and
\begin{subequations}
\label{eq:maximum_ss_FGFaGaFbGb}
\begin{align}
    F_a &= \begin{bmatrix}
        0 & 0_{n_y \times n_y} \\ 
        I_{n_y(n_a-1)} & 0
    \end{bmatrix} \in \mathbb{R}^{n_y n_a \times n_y n_a}
    &  &
    G_a = \begin{bmatrix}
        I_{n_y} \\ 0_{ n_y (n_a -1) \times n_y}
    \end{bmatrix}
    \in \mathbb{R}^{n_y n_a \times n_y}, 
    \label{eq:maximum_ss_Fa_Ga}
    \\
    F_b &= \begin{bmatrix}
        0 & 0_{n_u \times n_u} \\ 
        I_{n_u(n_b-2)} & 0
    \end{bmatrix} \in \mathbb{R}^{n_u(n_b-1) \times n_u(n_b -1)}
    & &
    G_b = \begin{bmatrix}
        I_{n_u} \\ 0_{n_u(n_b-2) \times n_u}
    \end{bmatrix}
    \in \mathbb{R}^{n_u(n_b-1) \times n_u},
    \label{eq:maximum_ss_Fb_Gb}
\end{align}
\end{subequations}
Even though realization \eqref{eq:maximum_state_space} can directly be constructed, it might contain more states than necessary, i.e., it might be a nonminimal realization. This can happen through two ways\footnote{Minimality in the Kalman sense is considered. If a realization is desired that also captures unreachable but observable autonomous behavior, then only the second mechanism is relevant.}: 1) it can include state directions that can never be reached, i.e., the state-space contains an unreachable subspace, and 2) it can include state directions that can never be observed, i.e., the state-space contains an unobservable subspace. 

The goal of this paper is to analyze observability and reachability properties \eqref{eq:maximum_state_space}, which are formally defined in the next Sections. Specifically, the goal is to provide numerical conditions in terms of the coefficient functions $A_i,B_i$ that imply observability and reachability.

Next to the general case in which $n_a>0,n_b>1$, also observability/reachability in the special cases that $n_a=0,n_b>1$ or $n_a>0,n_b=1$ are analyzed in this paper. In these cases, a state-space realization of \eqref{eq:LPV_IO_DT} is obtained by deleting the irrelevant blocks in \eqref{eq:maximum_state_space}. Specifically, in case $n_a=0$, \eqref{eq:LPV_IO_DT} reduces to an LPV finite-impulse response (FIR), for which a state-space realization of \eqref{eq:LPV_IO_DT} is given by
\begin{align} \label{eq:maximum_state_space_FIR}
    \bar{u}_{k+1} = F_b \bar{u}_k + G_b u_k  & & y_k = B(p_k) \bar{u}_k + B_0(p_k) u_k.
\end{align}
In case $n_b = 1$, IO representation \eqref{eq:LPV_IO_DT} reduces to an inverse LPV-FIR filter. A state-space realization of \eqref{eq:LPV_IO_DT} in this setting is given by
\begin{align} \label{eq:maximum_state_space_invFIR}
    \bar{y}_{k+1} = (F_a - G_a A(p_k)) \bar{y}_k + G_a B_0(p_k) u_k & & y_k = -A(p_k) \bar{y}_k + B_0(p_k) u_k,
\end{align}
with $A$ and $F_a,G_a$ as in \eqref{eq:maximum_ss_A} and \eqref{eq:maximum_ss_Fa_Ga}. 

\section{Reachability}
\label{sec:reachability}
In this section, a condition for reachability of the realization \eqref{eq:maximum_state_space} is derived in terms of the coefficient functions $A_i,B_i$, constituting contribution C1. 

Intuitively, reachability states that any point in the state space can be reached by an appropriate input. It is formally defined in the LPV case as follows \cite{Silverman1967, Gohberg1992, Toth2010}.

\begin{definition}
\label{def:reachability_DT}
    State-space realization \eqref{eq:maximum_state_space} is said to be structurally $k$-reachable if there exists a scheduling signal $p\in\signalspaceDT$ such that for any initial state $x_{k_1} \in \Xee$ at an arbitrary time $k_1$ and any final state $x_{k_2}\in\Xee$, $x_{k_1}$ can be steered to $x_{k_2}$ in $k$ steps. It is said to be completely $k$-reachable if this steering of the state is possible for all $p\in\signalspaceDT$.
\end{definition}
This definition of reachability mostly aligns with the LTI definition \cite[Section 6.1]{Chen1999}, but differs in two ways. First, Definition \ref{def:reachability_DT} discerns between structural and complete reachability. This stems from the fact that reachability might be lost for some scheduling signals $p$, e.g., through a pole-zero cancellation at a specific point $\bar{p}\in\Pee$. Structural reachability allows for such situations, whereas complete reachability does not. 

Second, Definition \ref{def:reachability_DT} defines reachability inherently with respect to a time interval. This again stems from the difference in behavior for different $p$: whereas all states might be reachable in $k=n_x$ steps for some $p$, they might only be reachable after, e.g., $k=10n_x$ steps for a different $p$. Structural reachability is a prerequisite for minimality of a realization \cite{Gohberg1992}: if all states are reachable for some $p$, these state directions contribute to the dynamics and cannot be projected out to obtain a smaller state dimension.

First, reachability of the state-space realization is investigated in the general case that $n_a > 0, n_b > 1$, i.e., \eqref{eq:maximum_state_space} is considered. The special cases $n_a = 0, n_b>1$ (FIR) or $n_a > 0, n_b =1$ (inverse FIR) are considered after. The next lemma defines a condition to verify structural $k$-reachability in terms of the $k$-step reachability matrix $\Rcal_k$.

\begin{lemma}
\label{lem:reachability_matrix}
    State-space realization \eqref{eq:maximum_state_space} is structurally k-reachable if and only if $\exists p\in \signalspaceDT$ such that $\rank\ \Rcal_k = n_x$ with
    \begin{equation}
        \Rcal_{k}(p_0,\ldots,p_{k-1}) = 
        \begin{bmatrix}
            \prod_{j=1}^{k-1} \Fcal(p_j) \Gcal(p_0) & \prod_{j=2}^{k-1} \Fcal(p_j) \Gcal(p_1) & \hdots & \Fcal(p_{k-1}) \Gcal(p_{k-2}) & \Gcal(p_{k-1})
        \end{bmatrix}
        \in \mathbb{R}^{n_x \times n_u k}, \label{eq:reach_matrix_generic}%
    \end{equation}%
    with $\Fcal,\Gcal$ as in \eqref{eq:maximum_ss_FcalGcal}.
\end{lemma}%
\begin{proof}
The proof follows from standard reachability arguments and is given here for completeness. First, note that $p$ is completely free such that $k_1,k_2$ in Definition \ref{def:reachability_DT} can be chosen as $k_1=0$ and $k_2=k$ without loss of generality. Then the evolution of $x_k$ in \eqref{eq:maximum_state_space} starting from $x_{k_1}  = x_0$ can be written as
    \begin{equation}
        x_k = \prod_{j=0}^{k-1} \Fcal(p_j) x_0 + \Rcal_k(p_0,\ldots,p_{k-1}) U_k,
    \end{equation}
    with $U_k = \begin{bmatrix} u_0^\T & u_1^\T & \ldots & u_{k-1}^\T \end{bmatrix}^\T$. Then, $x_0$ can be steered to any $x_k\in\Ree^{n_x}$ in $k$ steps if and only if $\im\ \Rcal_k = \Ree^{n_x}$, i.e., if and only if $\rank\ \Rcal_k = n_x$. A possible input that achieves this is $U_k = \Rcal_k^{\dag}(p_0,\ldots,p_{k-1}) \big(x_k - \prod_{j=0}^{k-1} A(p_j) x_0\big)$.
\end{proof}
Next, a sufficient condition for full rank of $\Rcal_{n_x}$, i.e., reachability in $n_x$ steps, is derived in terms of conditions on the coefficient functions $A_1,\ldots,A_{n_a}$ and $B_1,\ldots,B_{n_b}$ through considering \eqref{eq:maximum_state_space} for constant scheduling trajectories, i.e., $p_k = \bar{p} \ \forall k \in \Zee$ with $\bar{p}\in\Pee$. 

\begin{theorem}
    \label{th:DT_controllability}
    State-space realization \eqref{eq:maximum_state_space} with $n_a > 0, n_b > 1$ is structurally $n_x$-reachable if there exists a constant scheduling signal $p_k = \bar{p} \in \mathbb{P} \ \forall k \in \Zee$ such that
    \begin{subequations}
    \begin{align}
        &\textnormal{\rank} \begin{bmatrix} I + \sum_{i=1}^{n_a} \sigma^{-i}A_i(\bar{p}) & \sum_{i=0}^{n_b-1} \sigma^{-i} B_i(\bar{p}) \end{bmatrix} = n_y \ \forall \sigma \in \Cee \backslash \{ 0 \}, \label{eq:coprime_theorem} \\
        &\textnormal{\rank} \begin{bmatrix} -A_{n_a}(\bar{p}) & B_{n_b-1}(\bar{p}) \end{bmatrix} = n_y. \label{eq:last_coefficient_nullspace}
    \end{align}%
    \end{subequations}%
\end{theorem}
\begin{proof}
    The proof is based on the equivalence between the Popov-Belevitch-Hautus test for reachability and full rank of the associated reachability matrix for a constant scheduling signal \cite{hespanha2018linear}, i.e., in the LTI case, and proceeds similarly to the key reachability lemma \cite[Lemma 3.4.7]{Goodwin1984}. Specifically, define 
    \begin{equation}
        H(\sigma; \bar{p}) = \left[\begin{array}{c|c} \Fcal(\bar{p}) - \sigma I & \Gcal(\bar{p}) \end{array}\right] \in \Cee^{n_x \times (n_x + n_u)},
    \end{equation}
    with $\sigma\in\Cee$. Then $\rank\ H(\sigma; \bar{p}) = n_x \ \forall \sigma \in \mathbb{C}$ if and only if $\rank\ \Rcal_{n_x}(\bar{p},\ldots,\bar{p}) = n_x$ \cite{hespanha2018linear}, such that by Lemma \ref{lem:reachability_matrix}, realization \eqref{eq:maximum_state_space} is structurally $n_x$-reachable if $\rank\ H(\sigma; \bar{p}) = n_x \ \forall \sigma \in \mathbb{C}$. The remainder of this proof shows that $\rank\ H(\sigma; \bar{p}) = n_x \ \forall \sigma \in \mathbb{C}$ if and only if the conditions of Theorem \ref{th:DT_controllability} hold. 
    
    To show this, first substitute $\Fcal(\bar{p}),\Gcal(\bar{p})$ as in \eqref{eq:maximum_ss_ABCD} into $H(\sigma; \bar{p})$ to obtain
    \begin{equation}\label{eq:H_sigma_pbar_full}
    \resizebox{\textwidth}{!}{$
        H(\sigma;\bar{p}) \! = \! \left[
        \begin{array}{ccccc|ccccc|c}
        \!-A_1(\bar{p}) \!-\!\sigma I \! & \! -A_2(\bar{p}) \! & \! \hdots \! & \! -A_{n_a-1}(\bar{p}) \! & \! -A_{n_a}(\bar{p}) \! &  \!B_1(\bar{p}) \! & \! B_2(\bar{p}) \! & \! \hdots \! & \! B_{n_b-2}(\bar{p}) \! & \! B_{n_b-1}(\bar{p}) \! & \! B_0(\bar{p}) \! \\
        I & -\sigma I  & & 0 & 0 & 0 & 0 & \hdots & 0 & 0 & 0 \\
         & I & \hspace{0ex}\raisebox{2pt}{\rotatebox{15}{$\ddots$}} & & 0 & 0 & 0 & \hdots & 0 & 0 & 0 \\
        & & \hspace{0ex}\raisebox{2pt}{\rotatebox{15}{$\ddots$}}  & -\sigma I & \vdots & \vdots & \vdots & & \vdots & \vdots & \vdots \\
        0 &  & & I & -\sigma I & 0 & 0 & \hdots & 0 & 0 & 0\\
        \hline 
        0 & 0 & \hdots & 0 & 0 & -\sigma I &  & & 0  & 0 & I\\
        0 & 0 & \hdots & 0 & 0 & I & -\sigma I &  & & 0 & 0 \\
        0 & 0 & \hdots & 0 & 0 &  & I & \hspace{0ex}\raisebox{2pt}{\rotatebox{15}{$\ddots$}} &  & \vdots & \vdots \\
        \vdots & \vdots & & \vdots & \vdots & & & \hspace{0ex}\raisebox{2pt}{\rotatebox{15}{$\ddots$}} & -\sigma I & 0 & 0 \\
        0 & 0 & \hdots & 0 & 0 & 0 & & & I & -\sigma I & 0
        \end{array}
        \right]\!\!.
        $}
    \end{equation}
    Now consider that $\sigma \neq 0$ such that $\sigma^{-1}$ exists. Then define the matrices $T_{y},T_{u}$ representing elementary column operations as 
    \begin{subequations}
    \begin{align}
        T_{y} &=   \begin{bmatrix} I & & & & \\
                                  & \raisebox{0pt}{\rotatebox{10}{$\ddots$}} & & & \\
                                  & & I & & \\
                                  & & & I &  \\
                                  & & & \sigma^{-1} I & I \end{bmatrix}
                    \begin{bmatrix} I & & & & \\
                                  & \raisebox{0pt}{\rotatebox{10}{$\ddots$}} & & & \\
                                  & & I &   & \\
                                  & & \sigma^{-1} I& I & \\
                                  & & & & I \end{bmatrix}
                    \hdots
                    \begin{bmatrix} I & & & & \\
                                  \sigma^{-1} I& I & & & \\
                                  & & \raisebox{0pt}{\rotatebox{10}{$\ddots$}} &   & \\
                                  & & & I & \\
                                  & & & & I \end{bmatrix}
                    \in\Ree^{n_y n_a \times n_y n_a} \\
        T_{u} &=   \begin{bmatrix} I \vspace{-3pt} & & & & \\
                                  & \raisebox{0pt}{\rotatebox{10}{$\ddots$}} \hspace{-5pt} & & & \\
                                  & & \! I & & \\
                                  & & \sigma^{-1} I & I &  \\
                                  & & &  & I \end{bmatrix}
                    \hdots
                    \begin{bmatrix} I & & & & \\
                                  \sigma^{-1} I& I & & & \\
                                  & & \raisebox{0pt}{\rotatebox{10}{$\ddots$}} &   & \\
                                  & & & I & \\
                                  & & & & I \end{bmatrix} 
                    \begin{bmatrix} I & & & & \\
                                  & I & & & \\
                                  & & \raisebox{0pt}{\rotatebox{10}{$\ddots$}} &   & \\
                                  & & & I & \\
                                  & & & \sigma^{-1} I & I \end{bmatrix} \in \Ree^{(n_b+1)n_u \times (n_b+1)n_u}, \hspace{-10pt}
    \end{align}
    \end{subequations}
    such that
    \begin{align}
        &H(\sigma;\bar{p}) \begin{bmatrix}
            T_{y} & 0 \\
            0 & T_{u}
        \end{bmatrix}
        = \hat{H}(\sigma;\bar{p}) =   \label{eq:Hhat} \\
    &\resizebox{\textwidth}{!}{$
        \left[
        \begin{array}{ccccc|ccccc|c}
        \!-A_1(\bar{p})\! -\!\sigma I \!-\! \sum\limits_{i=2}^{n_a} \sigma^{-i+1} A_{i}(\bar{p}) \! & \! \ast \! & \! \hdots \! & \! \ast \! & \! \ast \! & \! \ast \! & \! \ast \! &\! \hdots \!& \! \ast \! & \! \ast \! & \! B_0(\bar{p}) \! +\! \sum\limits_{i=1}^{n_b-1} \sigma^{-i} B_i(\bar{p}) \\
        0 & \! -\sigma I \!  & & & 0 & 0 & 0 & \hdots & 0 & 0 & 0 \\
        0  &  & \raisebox{0pt}{\rotatebox{10}{$\ddots$}} & & & 0 & 0 & \hdots & 0 & 0 & 0 \\
        \vdots & & & \! -\sigma I \! & & \vdots & \vdots & & \vdots & \vdots & \vdots \\
        0 & 0 & &  & -\sigma I & 0 & 0 & \hdots & 0 & 0 & 0\\
        \hline 
        0 & 0 & \hdots & 0 & 0 & \!\!-\sigma I\!\! & & & & 0 & 0\\
        0 & 0 & \hdots & 0 & 0 & & \!\!-\sigma I\!\! &  & & & 0\\
        0 & 0 & \hdots & 0 & 0 &  & & \raisebox{0pt}{\rotatebox{10}{$\ddots$}} &  & & \vdots \\
        \vdots & \vdots & & \vdots & \vdots & & & & \!\!-\sigma I\!\! & & 0 \\
        0 & 0 & \hdots & 0 & 0 & 0 & & & & \!\!-\sigma I \!\!& 0
        \end{array}
        \right]\!,$} \nonumber
    \end{align}
    where $\ast$ indicates elements not of interest as they can be zeroed out using elementary row operations ($\sigma\neq0$). Note that $T_{y},T_{u}$ are full rank, such that $\rank\ H(\sigma;\bar{p}) = \rank\ \hat{H}(\sigma;\bar{p})$. Now $\hat{H}(\sigma;\bar{p})$ is upper block-triangular, and thus has full rank $n_x$ if and only if its block of rows has full rank, i.e., if and only if
    \begin{equation}
        \rank \begin{bmatrix} -\sigma I - A_1(\bar{p}) - \sum_{i=2}^{n_a} \sigma^{-i+1}A_i(\bar{p}) & \ \ \sum_{i=0}^{n_b-1} \sigma^{-i} B_i(\bar{p}) \end{bmatrix} = n_y, \label{eq:coprime_proof}
    \end{equation}
    which is equivalent to \eqref{eq:coprime_theorem} after a rank-preserving post-multiplication with $\textrm{blkdiag}(-\sigma^{-1}I, I)$.
    

    Turning attention to \eqref{eq:last_coefficient_nullspace}, consider that $\sigma = 0$ and define 
    \begin{equation}
        T = \left[\begin{array}{ccccccccccc} I_{n_y} & A_1(\bar{p}) & A_2(\bar{p}) & \hdots & A_{n_a-1}(\bar{p}) & -B_0(\bar{p}) & -B_1(\bar{p}) & -B_2(\bar{p}) & \hdots & -B_{n_b-2}(\bar{p}) & 0 \\ 0 & \multicolumn{10}{c}{I_{n_x-n_y}} \end{array}\right] \in \Ree^{n_x\times n_x}.
    \end{equation} 
    Then pre-multiplying $H(0;\bar{p})$ with $T$ yields
    \begin{equation}
        \bar{H}(0;\bar{p}) = T H(0;\bar{p}) = 
        \left[
        \begin{array}{ccccc|ccccc|c}
        0 & 0 & \hdots & 0 & -A_{n_a}(\bar{p}) & 0 & 0 & \hdots & 0 & B_{n_b-1}(\bar{p}) & 0 \\
        I & 0 & & 0 & 0 & 0 & 0 & \hdots & 0 & 0 & 0 \\
         & I & & & 0 & 0 & 0 & \hdots & 0 & 0 & 0 \\
        & & \ddots  & & \vdots & \vdots & \vdots & & \vdots & \vdots & \vdots \\
        0 &  & & I & 0 & 0 & 0 & \hdots & 0 & 0 & 0\\
        \hline 
        0 & 0 & \hdots & 0 & 0 & 0 & 0 & \hdots & 0 & 0 & I\\
        0 & 0 & \hdots & 0 & 0 & I & &  & 0 & 0 & 0 \\
        0 & 0 & \hdots & 0 & 0 &  & I & &  & 0 & 0 \\
        \vdots & \vdots & & \vdots & \vdots & & & \ddots & & \vdots & \vdots \\
        0 & 0 & \hdots & 0 & 0 & 0 & & & I & 0 & 0
        \end{array}
        \right],
        \label{eq:Hbar}
    \end{equation}
    in which $\rank\ \bar{H}(0;\bar{p}) = \rank\ H(0;\bar{p})$ as $T$ is square and full rank.
    By its structure, $\rank\ \bar{H}(0;\bar{p}) = n_x$ and only if its first row is full rank, i.e., $\rank \ \bar{H}(0;\bar{p}) = n_x$ if and only if \eqref{eq:last_coefficient_nullspace}, completing the proof. 
\end{proof}
Theorem \ref{th:DT_controllability} states that if coefficient functions $A_i,B_i$ satisfy \eqref{eq:coprime_theorem}-\eqref{eq:last_coefficient_nullspace} for some $\bar{p}\in\Pee$, then realization \eqref{eq:maximum_state_space} is structurally $n_x$-reachable. It can be understood by noting that conditions \eqref{eq:coprime_theorem}-\eqref{eq:last_coefficient_nullspace} are necessary and sufficient for reachability of the frozen LTI behavior of \eqref{eq:maximum_state_space} at a point $\bar{p}$, such that \eqref{eq:maximum_state_space} is structurally $n_x$-reachable if the frozen LTI behavior is reachable for some point $\bar{p}$. Condition \eqref{eq:coprime_theorem} states that there should be at least one constant value for the scheduling signal $\bar{p}$ for which no pole-zero cancellations occur, i.e., it represents a coprime condition \cite[Chapter 7]{Chen1999} over $\bar{p}\in\Pee$. In the case of single-input single-output (SISO) systems, this condition merely states that the polynomials $P(\sigma; \bar{p}) = I+\sum_{i=1}^{n_a} A_i(\bar{p}) \sigma^{-i}$ and $Q(\sigma; \bar{p}) = \sum_{i=0}^{n_b-1} B_i(\bar{p}) \sigma^{-i}$ should not have a common root for all $\bar{p}\in\Pee$. In the multivariable case, directionality of these roots is taken into account. Condition \eqref{eq:last_coefficient_nullspace} states that the order of the IO representation \eqref{eq:LPV_IO_DT} should not be unnecessarily high, i.e., it can be interpreted as a well-posedness condition. Specifically, in the SISO case, it states that either $A_{n_a}(\bar{p}) \neq 0$ or $B_{n_b-1}(\bar{p}) \neq 0$. By choosing the model order appropriately, i.e., by not incorporating a coefficient that is zero for all $\bar{p}\in\Pee$, this condition is automatically satisfied. In the multivariable case, the directionality of $A_{n_a}(\bar{p})$ and $B_{n_b-1}(\bar{p})$ is taken into account.

Last note that Theorem \ref{th:DT_controllability} only considers constant scheduling trajectories, and is thus only a sufficient condition for structural reachability of \eqref{eq:maximum_state_space}. In other words, $\Rcal_k$ can be full rank for a varying scheduling signal, even if it is not for any constant scheduling $\bar{p}\in\Pee$.

Condition \eqref{eq:coprime_theorem} imposes a condition for all $\sigma \in \Cee \backslash \{0\}$. However, since $\rank\ P(\sigma;\bar{p}) = n_y$ for $\sigma \in \Cee$ that are not roots of $P(\sigma;\bar{p})$, this condition only has to be verified at these roots. Thus, a practical way to evaluate this condition for a given LPV model is to evaluate the coefficient functions for a grid of $\bar{p}\in\Pee$, calculating the roots of $P(\sigma;\bar{p})$ over this grid, and evaluating \eqref{eq:coprime_theorem} at the roots of $P(\sigma)$, i.e., checking for pole-zero cancellations over a grid of $\bar{p}\in\Pee$.

Lemma \ref{lem:reachability_matrix} and Theorem \ref{th:DT_controllability} can straightforwardly be specialized to the special cases $n_b=1$ (inverse LPV-FIR) with realization \eqref{eq:maximum_state_space_invFIR} or $n_a=0$ (LPV-FIR) with realization \eqref{eq:maximum_state_space_FIR} as follows. 
\begin{corollary}\label{cor:reachability_invFIR}
    State-space realization \eqref{eq:maximum_state_space_invFIR} is structurally $n_x$-reachable if there exists a constant scheduling signal $p_k = \bar{p} \in\Pee\ \forall k\in\Zee$ such that $\rank\ B_0(\bar{p}) = n_y$.
\end{corollary}
\begin{proof}
    The proof is analogous to the proof of Theorem \ref{th:DT_controllability} and links the rank of $\Rcal_k$ in \eqref{eq:reach_matrix_generic} for a constant scheduling $\bar{p}$ to the rank of the PBH matrix pencil $H(\sigma; \bar{p}) \in \Ree^{n_x \times n_x + n_u}$. Specifically, for this setting, it holds that $H(\sigma;\bar{p}) = \left[\begin{array}{c|c} F_a - G_a A(p_k) & G_a B_0 (p_k) \end{array}\right]$, or 
    \begin{equation}
        H(\sigma;\bar{p}) \! = \! \left[
        \begin{array}{ccccc|c}
        \!-A_1(\bar{p}) \!-\!\sigma I \! & \! -A_2(\bar{p}) \! & \! \hdots \! & \! -A_{n_a-1}(\bar{p}) \! & \! -A_{n_a}(\bar{p}) \! &  \! B_0(\bar{p}) \! \\
        I & -\sigma I  & & 0 & 0 & 0 \\
         & I & \raisebox{0pt}{\rotatebox{10}{$\ddots$}} & & \vdots & \vdots \\
        & & \raisebox{0pt}{\rotatebox{10}{$\ddots$}}  & -\sigma I & 0 & 0 \\
        0 &  & & I & -\sigma I & 0 
        \end{array}
        \right]
    \end{equation}
    Carrying out the same steps as in the proof of Theorem \ref{th:DT_controllability} yields that $\rank\ H(\sigma; \bar{p}) = n_x \ \forall \sigma\in\Cee$ if and only if
    \begin{align}
        \rank \begin{bmatrix} I + \sum_{i=1}^{n_a} \sigma^{-1} A_i(\bar{p}) & B_0(\bar{p}) \end{bmatrix} = n_y \ \forall \sigma \in \Cee \backslash \{0\} & & \textnormal{\rank}\ B_0(\bar{p}) = n_y.
    \end{align}
    Now, the first condition is implied by the second, such that $\rank\ H(\sigma; \bar{p}) = n_x \ \forall \sigma\in\Cee$ if and only if $\rank\ B_0(\bar{p}) = n_y$. Then, by standard LTI arguments, $\rank\ \Rcal_k = n_x$ for some constant scheduling $\bar{p}$ if and only if $\rank\ H(\sigma; \bar{p}) = n_x \ \forall \sigma\in\Cee$, i.e., \eqref{eq:maximum_state_space_invFIR} is structurally reachable if $\rank\ B_0(\bar{p}) = n_y$ for some $\bar{p}\in\Pee$.
\end{proof}
\begin{corollary}\label{cor:reachability_FIR}
    State-space realization \eqref{eq:maximum_state_space_FIR} is completely $n_x$-reachable.
\end{corollary}
\begin{proof}
    For realization \eqref{eq:maximum_state_space_FIR}, the state-transition equation $\bar{u}_{k+1} = F_b \bar{u}_k + G_b u_k$ is time-invariant, such that reachability is no longer dependent on the scheduling signal $p$. Its reachability matrix $R_{n_x}$ satisfies $R_{n_x} = I$ for any $p$. Thus, \eqref{eq:maximum_state_space_FIR} is completely $n_x$-reachable.
\end{proof}
Corollary \ref{cor:reachability_FIR} can be intuitively understood by noting that the states of \eqref{eq:maximum_state_space_FIR} are simply shifted versions of $u_k$, which can be directly set by the input $u_k$.  

Taken together, Theorem \ref{th:DT_controllability} and Corollaries \ref{cor:reachability_invFIR} and \ref{cor:reachability_FIR} show that the state-space resulting from the direct state construction \eqref{eq:maximum_ss_state_def} is reachable under coprimeness and well-posedness conditions.

\section{Observability and Reconstructability}\label{sec:observability}
In this section, observability and reconstructability properties of \eqref{eq:maximum_state_space} are  derived, constituting contribution C2. Specifically, it is shown that for state-space realization \eqref{eq:maximum_state_space}, there exist initial states of past inputs and outputs of which the effect cannot be observed in the output, i.e., \eqref{eq:maximum_state_space} has an unobservable subspace and is thus not minimal. Even though this unobservable subspace is present, it is shown that for any coefficient functions $A_i$ and $B_i$, the states in this unobservable subspace decay to 0 in at most $\max(n_a,n_b-1)$ time steps, such that realization \eqref{eq:maximum_state_space} can never have unstable unobservable modes. To formalize these claims, consider the definition of observability and reconstructability in the LPV case \cite{Silverman1967, Gohberg1992, Toth2010}.

\begin{definition}
\label{def:obsv_DT}
   State-space realization \eqref{eq:maximum_state_space} is said to be structurally observable in $k\in\Nee$ steps if there exists a scheduling signal $p\in\signalspaceDT$ such that for any initial state $x_{k_1}\in\Xee$ at an arbitrary time $k_1\in\Zee$ and any input signal $u\in\signalspaceDT$, $x_{k_1}$ can be uniquely reconstructed from $\{ u_{j}, y_{j} \}_{j=k_1}^{k_1+k-1}$. It is said to be completely $k$-observable if the reconstruction of $x_{k_1}$ is possible for all $p\in\signalspaceDT$.
\end{definition}
Structural observability means that for at least one scheduling signal $p$, any state at time $k_1$ can be inferred from input-output data \textit{after} $k_1$. In contrast, complete observability requires that this is possible for all $p$. Structural observability strongly relates to minimality of the state space realization. Specifically, if a subspace of $\Xee$ is observable for some $p$, i.e., if this subspace is structurally observable, this subspace contributes to the input-output behavior of \eqref{eq:maximum_state_space}. Next to observability, the weaker notion of structural $k$-reconstructability is defined \cite{Silverman1967, Gohberg1992, Toth2010}.

\begin{definition}
\label{def:recon_DT}
    State-space realization \eqref{eq:maximum_state_space} is said to be structurally reconstructible in $k\in\Nee$ steps if there exists a scheduling signal $p\in\signalspaceDT$ such that for any initial state $x_{k_1}\in\Xee$ at an arbitrary time $k_1\in\Zee$ and any input signal $u\in\signalspaceDT$, $x_{k_1+k}$ can be uniquely reconstructed from $\{ u_{j}, y_{j} \}_{j=k_1}^{k_1+k-1}$. It is said to be completely $k$-reconstructible if the reconstruction of $x_{k_1+k}$ is possible for all $p\in\signalspaceDT$.
\end{definition}
Structural reconstructability means that for at least one scheduling signal $p$, the state at time $k_1+k$ can be inferred from input-output data \textit{before} $k_1+k$. In contrast to observability, only the part of $x_{k_1}$ that does not decay to 0 in $k$ steps is required to be reconstructible. Consequently, structural $k$-reconstructability allows for an unobservable subspace as long as states in this subspace decay to $0$ in $k$ steps, and is thus a weaker notion than observability. Last, note that detectability follows as reconstructability for $k\rightarrow\infty$. 

As in the reachability case, first observability and reconstructability are investigated in the general case that $n_ a> 0,n_b>1$, i.e., \eqref{eq:maximum_state_space} is considered. Afterwards, the cases in which $n_b=1$ or $n_a =0$ are considered, described by \eqref{eq:maximum_state_space_invFIR} and \eqref{eq:maximum_state_space_FIR} respectively. The next lemma provides a condition to verify observability and reconstructability in terms of the rank of the $k$-step observability matrix $\Ocal_k$. 


\begin{lemma}
\label{lem:observability_matrix}
    State-space realization \eqref{eq:maximum_state_space} is structurally k-observable if and only if $\exists p\in \signalspaceDT$ such that $\ker\ \Ocal_k = \emptyset$ with
    \begin{equation}
        \Ocal_{k} = 
        \begin{bmatrix}
            \Hcal(p_{0}) \\
            \Hcal(p_{1}) \Fcal(p_{0}) \\
            \Hcal(p_{2}) \Fcal(p_{1}) \Fcal(p_{0}) \\
            \vdots \\
            \Hcal(p_{k-1}) \prod_{j=0}^{k-2} \Fcal(p_{j})
        \end{bmatrix}
        \in \mathbb{R}^{n_y k \times n_x}
        \label{eq:obsv_matrix_generic}
    \end{equation}
    It is completely $k$-reconstructible if and only if $\ker\ \Ocal_k \subseteq \ker\prod_{j=0}^{k-1} \Fcal(p_{j})$ for all scheduling signals $p\in \signalspaceDT$.
\end{lemma}
\begin{proof}
    The proof follows from standard observability arguments \cite{Gohberg1992} and is given here for completeness. Throughout, $k_1$ in Definitions \ref{def:obsv_DT} and \ref{def:recon_DT} is chosen as $k_1 = 0$ without loss of generality, as the moment of zero time can be chosen arbitrary. Then, the response $y_j$ with $j$ ranging from $k_1=0$ to $k-1$ from initial state $x_0$ is given by
    \begin{equation}\label{eq:DT_response}
    \begin{aligned}
        \underbrace{\begin{bmatrix}
            y_0 \\ y_1 \\ y_{2} \\ \vdots \\ y_{k-1}
        \end{bmatrix}}_{y_{[0:k-1]}}
        &=
        \Ocal_k x_0
        -
        \underbrace{\begin{bmatrix}
        \Jcal(p_0) & & & 0\\
        \Hcal(p_1) \Gcal(p_0) & \Jcal(p_1) & & \\
        \Hcal(p_2) \Fcal(p_1) \Gcal(p_0) & \Hcal(p_2) \Gcal(p_1) & &  \\
        \vdots & \vdots & \raisebox{0pt}{\rotatebox{22}{$\ddots$}}\\
        \Hcal(p_{k-1}) \prod_{j=1}^{k-2} \Fcal(p_j) \Gcal(p_0) & \Hcal(p_{k-1}) \prod_{j=2}^{k-2} \Fcal(p_j) \Gcal(p_1) & \hdots & \Jcal(p_{k-1})
        \end{bmatrix}}_{\Gamma}
        \underbrace{\begin{bmatrix}
            u_0 \\ u_1 \\ u_2 \\ \vdots \\ u_{k-1}
        \end{bmatrix}}_{u_{[0:k-1]}}.
    \end{aligned}
    \end{equation}

    Then, for the observability case, $x_0$ can be uniquely determined from $\{u_j, y_j\}_{j=0}^{k-1}$ by $\Ocal_k x_0 = (y_{[0:k-1]} - \Gamma u_{[0:k-1]})$ if and only if $\ker\ \Ocal_k = \emptyset$, i.e., realization \eqref{eq:maximum_state_space} is structurally $k$-observable if and only if $\exists p\in \signalspaceDT$ such that $\ker\ \Ocal_k = \emptyset$. 
    
    Next, for $k$-reconstructability case, note that $x_{k+k_1} = x_{k}$ is given by
    \begin{equation}
        x_k = \prod_{j=0}^{k-1} \Fcal(p_j) x_0 + \sum_{i=0}^{k-1} \prod_{j=i+1}^{k-1} \Fcal(p_j) \Gcal(p_i) u_i.
        \label{eq:x_k_pred}
    \end{equation}
    Then any reconstruction of $x_0$ from \eqref{eq:DT_response} in case $\ker\ \Ocal_k \neq \emptyset$ yields that the reconstruction $\hat{x}_0$ satisfies $\hat{x}_0 = x_0 + \delta x $ for some $\delta x \in \ker\ \Ocal_k$, as these $\delta x$ cannot be inferred from the output $y_0,\ldots,y_{k-1}$.
    Substituting this estimate $\hat{x}_0$ for $x_0$ in \eqref{eq:x_k_pred} yields an estimate $\hat{x}_k$, which correctly reconstructs $x_k$ if and only if $\delta x \in \ker\prod_{j=0}^{k-1} \Fcal(p_j)$, i.e., state-space realization \eqref{eq:maximum_state_space} is completely $k$-reconstructible if and only if $\ker\ \Ocal_k \subseteq \ker \prod_{j=0}^{k-1} \Fcal(p_j) \ \forall p \in \signalspaceDT$.
\end{proof}

Lemma \ref{lem:observability_matrix} can be understood by noting that the kernel of the observability matrix $\Ocal_k$ directly gives the directions that cannot be observed from the output. Given this characterization of observability and reconstructability in terms of the kernel of $\Ocal_k$, first it is shown that \eqref{eq:maximum_state_space} is completely reconstructible in $\max(n_a,n_b-1)$ steps.
\begin{theorem}\label{th:reconstructability}
    State-space realization \eqref{eq:maximum_state_space} is completely $max(n_a,n_b-1)$-reconstructible.
\end{theorem}
\begin{proof}
The proof is based on noting that any $x_k$ that does not produce an output at time $k$, i.e., $H(p_k) x_k =0$, also satisfies $x_{k+1} = \Fcal(p_k)x_k = (F+G\Hcal(p_k))x_k = F x_k$, i.e., such $x_k$ is only propogated in time by $F$. By the structure of $F$, any such $x_k$ must fade out in $\max(n_a,n_b-1)$ steps. Formally, set $k = \max(n_a,n_b-1)$ and consider $\Ocal_k$ for any scheduling signal $p$
, and consider any initial state $x_{0} \in\ker\ \Ocal_k$. For such an $x_0$, it is claimed that it holds that $\prod_{j=0}^{k-1}\Fcal(p_j) x_0 = F^{k} x_0$. To see this, note that for $x_0 \in \ker\ \Ocal_k$ it must hold that $\Hcal(p_0) x_0 = 0$ by the first rows of $\Ocal_k$, which implies that $\Fcal(p_0) x_0 = (F + G \Hcal(p_0)) x_0 = F x_0$. Similarly, by the second row of $\Ocal_k$, it holds that $\Hcal(p_1) \Fcal(p_0) x_0 = 0$, which implies that $\Fcal(p_1) \Fcal(p_0) x_0 = (F+G\Hcal(p_1)) F x_0 = F^2 x_0$. Continuing, it holds that $\prod_{j=0}^{k-1} \Fcal(p_j) x_0 = F^{k} x_0$. Now, for $F$ it holds that $F^{j} = 0$ for $j\geq k$, i.e., $F$ is a nilpotent matrix. Consequently, any $x_0 \in\ker\ \Ocal_k$ satisfies $\prod_{j=0}^{k-1} \Fcal(p_j) x_0 = 0$, or $\ker\ \Ocal_k \subseteq \ker\prod_{j=0}^{k-1} \Fcal(p_j)$ for $k = \max(n_a,n_b-1)$.
\end{proof}

Intuitively, it should come at no surprise that \eqref{eq:maximum_state_space} is completely $\max(n_a,n_b-1)$-reconstructible: the time segment $\{k,k_1,\ldots,k\!+\max(n_a,n_b\!-\!1)\!-\!1\}$ contains all inputs and outputs that constitute the state of \eqref{eq:maximum_state_space} at $k+\max(n_a,n_b-1)$, see the state definition \eqref{eq:maximum_ss_state_def}. Thus, the state at $k+\max(n_a,n_b-1)$ is trivially reconstructible. Theorem \ref{th:reconstructability} directly implies that state-space realization \eqref{eq:maximum_state_space} cannot have unstable unobservable states.

Even though \eqref{eq:maximum_state_space} is completely reconstructible, it is never observable. Specifically, to analyze observability of \eqref{eq:maximum_state_space} for a given set of coefficient functions, $\Ocal_k$ is characterized in terms of $A_i,B_i$. Using this characterization, it is shown that the kernel of $\Ocal_k$ is not empty for any $k\in\Nee$.

\begin{lemma}\label{lem:Ocalbar}
    Given any $k\in\Nee$, there exists a full rank matrix $T_k \in \Ree^{n_x \times n_x}$ such that the observability matrix $\Ocal_k$ can be written as $T_k \Ocal_k = \bar{\Ocal}_k$ with 
        \begin{equation} \label{eq:Ocalbar}
        \bar{\Ocal}_{k} = \begin{bmatrix}
        -A(p_0) & B(p_0) \\
        -A(p_1) F_a & B(p_1) F_b \\
        -A(p_2) F_a^2 & B(p_2) F_b^2 \\
        \vdots & \vdots \\
        -A(p_{k-1}) F_a^{k-1} & B(p_{k-1}) F_b^{k-1} \\
    \end{bmatrix} \in \Ree^{n_y k \times n_x}.
    \end{equation}
    in which $F_a,F_b$ are given by \eqref{eq:maximum_ss_FGFaGaFbGb}.
\end{lemma}
\begin{proof}
    Define
    \begin{align}
        F &= \left[\begin{array}{c|c}
        F_a & 0 \\ \hline 0 & F_b
        \end{array}\right]
        \in \mathbb{R}^{n_x \times n_x} 
        & & 
        G = \left[\begin{array}{c}
            G_a \\ \hline 0
        \end{array}\right]
        \in \mathbb{R}^{n_x \times n_y}
        \label{eq:maximum_ss_F_G}
    \end{align}
    with $F_a,F_b,G_a$ as in \eqref{eq:maximum_ss_FGFaGaFbGb}. Then $\Fcal$ in \eqref{eq:maximum_ss_ABCD} can be written as $\Fcal(p_j) = F + G \Hcal(p_j)$. Now it is noted that the state-transition matrix $\Fcal$ is a function of the output matrix $\Hcal$, which allows for simplifying the observability matrix in \eqref{eq:obsv_matrix_generic}. Specifically, define $\Ncal_j = \Hcal(p_{j-1}) \prod_{i=0}^{j-2} \Fcal(p_i)$ for $j=1,\ldots,k$, i.e., $\Ncal_j$ is the $j^{\textnormal{th}}$ block of rows of \eqref{eq:obsv_matrix_generic}. With these definitions, it is claimed that it holds that
    \begin{equation}\label{eq:aux}
        \mathcal{N}_j =  \Hcal(p_{j-1}) F^{j-1} +  \Hcal(p_{j-1}) \sum_{i=0}^{j-2} F^{j-i}G \mathcal{N}_i.
    \end{equation}
    To see this intuitively, for $j=1,2,3$ it holds that
    \begin{equation}
    \begin{aligned}
        \mathcal{N}_1 &= \Hcal(p_0) \\
        \mathcal{N}_2 &= \Hcal(p_1) \Fcal(p_0) = \Hcal(p_1) (F + G \Hcal(p_0)) = \Hcal(p_1) F + \Hcal(p_1) G \underbrace{\Hcal(p_0)}_{\mathcal{N}_1} \\
        \mathcal{N}_3 &= \Hcal(p_2) \Fcal(p_1) \Fcal(p_0) = \Hcal(p_2) (F + G \Hcal(p_1)) (F + G \Hcal(p_0)) \\
        &= \Hcal(p_2) (F^2 + G \underbrace{\Hcal(p_1) (F + G \Hcal(p_0))}_{\mathcal{N}_2} + F G \underbrace{\Hcal(p_0)}_{\mathcal{N}_1}). 
    \end{aligned}
    \end{equation}
    Formally, for arbitrary $j$ it holds that
    \begin{equation}
        \begin{aligned}
            \mathcal{N}_j &= \Hcal(p_{j-1}) \prod_{i=0}^{j-2} \Fcal(p_i) = \Hcal(p_{j-1}) (F + G \Hcal(p_{j-2})) \prod_{i=0}^{j-3} (F + G\Hcal(p_i)) \\
            &= \Hcal(p_{j-1}) G \Hcal(p_{j-2}) \prod_{i=0}^{j-3} (F + G\Hcal(p_i)) + \Hcal(p_{j-1}) F \prod_{i=0}^{j-3} (F + G\Hcal(p_i))
            \label{eq:N_K_proof}
        \end{aligned}
    \end{equation}
    where, for the first term, it holds by definition that $\mathcal{N}_{j-1}=\Hcal(p_{j-2}) \prod_{i=0}^{j-3} (F + G\Hcal(p_i))$, see above \eqref{eq:aux}. Applying a similar expansion to the second term results in 
    \begin{equation}
        \begin{aligned}
            \mathcal{N}_j &= \Hcal(p_{j-1}) G \mathcal{N}_{j-1} + \Hcal(p_{j-1}) F (F+GJ(p_{j-3})) \prod_{i=0}^{j-4} (F+G\Hcal(p_i)) \\
            &= \Hcal(p_{j-1}) G \mathcal{N}_{j-1} + \Hcal(p_{j-1}) F G \Hcal(p_{j-3}) \prod_{i=0}^{j-4} (F+G\Hcal(p_i)) + \Hcal(p_{j-1}) F^2 \prod_{i=0}^{j-4} (F+G\Hcal(p_i)),
        \end{aligned}
    \end{equation}
    where for the second term it holds that $\Hcal(p_{j-3}) \prod_{i=0}^{j-4} (F+G\Hcal(p_i)) = \mathcal{N}_{j-2}$. Continuing all the way gives
    \begin{align}
        \mathcal{N}_j &= \Hcal(p_{j-1}) G \mathcal{N}_{j-1} + \Hcal(p_{j-1}) F G \mathcal{N}_{j-2} + \Hcal(p_{j-1}) F^2 G \mathcal{N}_{j-3} + \ldots + \Hcal(p_{j-1}) F^{j-2} G \mathcal{N}_{1} + \Hcal(p_{j-1}) F^{j-1} \nonumber \\
        &= \Hcal(p_{j-1}) F^{j-1} + \Hcal(p_{j-1}) \sum_{i=1}^{j-1} F^{j-1-i}G \mathcal{N}_i.
        \label{eq:N_k_proof2}
    \end{align}
    Using this expansion for $\mathcal{N}_j$, the terms $\Ncal_1,\ldots,\Ncal_k$ satisfy
    \begin{equation}
        \underbrace{
        \begin{bmatrix}
            I & & & & &  0 \\
            -\Hcal(p_1) G & I & & & &  \\
            -\Hcal(p_2) F G & -\Hcal(p_2) G & I & & & \\
            \vdots & \vdots & & \raisebox{3pt}{\rotatebox{25}{$\ddots$}} & & \\
            -\Hcal(p_{k-2})F^{k-3}G & \!\! -\Hcal(p_{k-2})F^{k-4}G \!& \hdots & \!\!\!\!\!\!\!\!-\Hcal(p_{k-2})G & I &\\
            -\Hcal(p_{k-1})F^{k-2} G &\!\! -\Hcal(p_{k-1})F^{k-3} G \! & \hdots & \!\!\!-\Hcal(p_{k-1})FG & \! -\Hcal(p_{k-1})G \! & \! I  
        \end{bmatrix}
        }_{T_k}
        \underbrace{
        \begin{bmatrix}
            \mathcal{N}_1 \\
            \mathcal{N}_2 \\
            \mathcal{N}_3 \\
            \vdots \\
            \mathcal{N}_{k-1} \\
            \mathcal{N}_{k}
        \end{bmatrix}
        }_{\Ocal_k}
        \!=\! 
        \underbrace{
        \begin{bmatrix}
            \Hcal(p_0) \\
            \Hcal(p_1) F \\
            \Hcal(p_2) F^2 \\
            \vdots \\
            \Hcal(p_{k-2}) F^{k-2} \\
            \Hcal(p_{k-1}) F^{k-1}
        \end{bmatrix}
        }_{\bar{\Ocal}_k}
        \!.
    \end{equation}
    Now note that $T_k$ is lower block diagonal with $I$ on its main diagonal, i.e., it is unimodular and full rank. Consequently, it holds that $\rank\ \Ocal_k = \rank\ \bar{\Ocal}_k$ and $\ker\ \Ocal_k = \ker\ \bar{\Ocal}_k$, such that $\bar{\Ocal}_k$ can be equivalently considered. Using the partitioned forms $\Hcal(p_j) = \left[\begin{array}{c|c} -A(p_j) & B(p_j) \end{array}\right]$ and $F = \textrm{blkdiag}(F_a,F_b)$, see \eqref{eq:maximum_ss_ABCD} and \eqref{eq:maximum_ss_FGFaGaFbGb}, $\bar{\Ocal}_k$ can be written as \eqref{eq:Ocalbar}.   
\end{proof}
Even though above Lemma is only an algebraic manipulation of $\Ocal_k$, it enables analyzing $\ker\ \Ocal_k$ through $\ker\ \bar{\Ocal}_k$ as $T_k$ is full rank. This enables the following observability result.
\begin{theorem}\label{th:observability}
    For the state-space realization \eqref{eq:maximum_state_space} it holds that $\rank\ \Ocal_{k} < n_x$ for all $k\in\Nee$ and all $p\in\signalspaceDT$, i.e., \eqref{eq:maximum_state_space} is not observable. Furthermore, $\ker\ \Ocal_k$ is given by $\ker\ \bar{\Ocal}_k$ with $\bar{\Ocal}_k$ in \eqref{eq:Ocalbar}.
\end{theorem}
\begin{proof}
    The proof is based on analysis of $\bar{\Ocal}_k$ and nilpotency of shift matrices $F_a$,$F_b$. Specifically, first note that by Lemma \ref{lem:Ocalbar} and full rank of $T_k$, $\ker\ \Ocal_k = \ker\ \bar{\Ocal}_k$ and $\rank\ \Ocal_k = \rank\ \bar{\Ocal}_k$, i.e., $\bar{\Ocal}_k$ can be considered to derive observability properties of  \eqref{eq:maximum_state_space}. Then the rank of $\bar{\Ocal}_k$ is analyzed in 4 separate cases, which taken together imply $\rank\ \Ocal_{k} < n_x$ for all $k$ and all $p\in\signalspaceDT$.
    \begin{enumerate}[leftmargin=*]
        \item $k \leq n_a$: in this case $k n_y < n_x$, as $n_x = n_y n_a + n_u (n_b-1)$ and $n_b > 1$, i.e., $\bar{\Ocal}_k \in \Ree^{k n_y \times n_x}$ is wide and its rank trivially satisfies $\rank\ \bar{\Ocal}_k \leq k n_y < n_x$.
        \item $n_a \geq n_b -1$ and $k > n_a$: for $F_a$, it holds that $F_a^{k} \neq 0$ for $0 \leq k < n_a$ and $F_a^{k} = 0$ for $k \geq n_a$. Similarly, for $n_b$ it holds that $F_b^k \neq 0$ for $0\leq k < n_b-1$ and $F_b^{k} = 0$ for $k \geq n_b-1$. Thus $\bar{\Ocal}_k$ is given by
        \begin{equation}
            \bar{\Ocal}_k = 
            \left[\begin{array}{cc}
            -A(p_0) & B(p_0) \\
            \vdots & \vdots \\
            -A(p_{n_b-2}) F^{n_b-2} & B(p_{n_b-2}) F^{n_b-2}\\
            \hline
            -A(p_{n_b-1})F^{n_b-1} & 0 \\
            \vdots & \vdots \\
            -A(p_{n_a-1}) F^{n_a-1} & 0 \\
            \hline
            0_{n_y(k - n_a) \times n_y n_a} & 0_{n_y(k-n_a) \times n_u (n_b-1)}
            \end{array}\right],
        \end{equation}
        for $n_a > n_b-1$. In case $n_a = n_b-1$, the middle blocks are not present. In both cases, $\bar{\Ocal}_k$ only has $n_a n_y$ rows that are potentially nonzero depending on $p$, i.e., $\rank\ \bar{\Ocal}_k \leq n_a n_y < n_x$.
        \item $n_a < n_b -1$ and $k>n_b-1$ (and thus $k > n_a$): by the same nilpotency properties, $\bar{\Ocal}_k$ is given by
        \begin{equation}\label{eq:aux2}
            \bar{\Ocal}_k = 
            \left[\begin{array}{cc}
            -A(p_0) & B(p_0) \\
            \vdots & \vdots \\
            -A(p_{n_a-1}) F^{n_a-1} & B(p_{n_a-1}) F^{n_a-1}\\
            \hline
            0 & B(p_{n_a}) F^{n_a} \\
            \vdots & \vdots \\
            0 &  B(p_{n_b-2}) F^{n_b-2} \\
            \hline
            0_{n_y(k - n_b-1) \times n_y n_a} & 0_{n_y(k - n_b-1) \times n_u (n_b-1)}
            \end{array}\right] 
            =
            \left[\begin{array}{cc}
            \bar{\Ocal}_{[0,n_a-1],y} & \bar{\Ocal}_{[0,n_a-1],u} \\ \hline 0 & \bar{\Ocal}_{[n_a,n_b-2],u} \\ \hline 0 & 0
            \end{array}\right],
        \end{equation}
        with $\bar{\Ocal}_{[0,n_a-1],y} \in \Ree^{n_y n_a \times n_y n_a}$, $\bar{\Ocal}_{[0,n_a-1],u} \in\Ree^{n_y n_a \times n_y (n_b-1)}$ and $\bar{\Ocal}_{[n_a,n_b-2],u} \in \Ree^{n_y (n_b - 1 - n_a) \times n_u (n_b-1-n_a)}$. 
        Now, if $n_u \geq n_y$, then $\left[\begin{array}{cc} \bar{\Ocal}_{[0,n_a-1],y} & \bar{\Ocal}_{[0,n_a-1],u} \\ \hline 0 & \bar{\Ocal}_{[n_a,n_b-2],u}\end{array}\right]\in\Ree^{n_y (n_b-1) \times n_x}$ is wide since $n_x = n_a n_y + n_u (n_b-1) > n_y (n_b-1)$ for $n_u > n_y$ and $n_a > 0$. Consequently, $\rank\ \bar{\Ocal}_k \leq  n_y (n_b-1) < n_x$ as the remaining rows are 0. Alternatively, in case $n_u < n_y$, $\bar{\Ocal}_k$ is tall and it holds that
        \begin{equation}
            \dim (\ker \left[\begin{array}{cc} 0 & \bar{\Ocal}_{[n_a,n_b-2],u}\end{array}\right]^\perp) = n_u(n_b-1-n_a),
        \end{equation}
        such that the row space of $\bar{\Ocal}_k$ satisfies 
        \begin{equation}
        \begin{aligned}
            \dim ( (\ker\ \bar{\Ocal_k})^\perp) 
            \leq &
            \dim (\ker \left[\begin{array}{cc}
                \bar{\Ocal}_{[0,n_a-1],y} & \bar{\Ocal}_{[0,n_a-1],u}
            \end{array}\right]^\perp)
            + 
            \dim (\ker \left[\begin{array}{cc}
            0 & \bar{\Ocal}_{[n_a,n_b-2],u}
            \end{array}\right]^\perp) 
            \\
            & \leq n_y n_a + n_u(n_b-1-n_a) < n_x
        \end{aligned}
        \end{equation}
        where the last inequality holds as $n_a > 0$. Consequently, also for $n_u < n_y$, it holds that $\rank\ \bar{\Ocal}_k < n_x$.
            
        \item $n_a < n_b -1$ with $k>n_a$ but $k \ngtr n_b-1$: in this case, $\bar{\Ocal}_k$ is a submatrix of \eqref{eq:aux2}, such that directly  $\rank\ \bar{\Ocal}_k < n_x$.
    \end{enumerate}
    Summarizing, by cases 2-4, it holds that $\rank\ \bar{\Ocal}_k < n_x$ for $k > n_a$, and together with 1 thus $\rank\ \bar{\Ocal}_k < n_x$ for all $k>0$ and all $p\in\signalspaceDT$.
\end{proof}

Theorem \ref{th:observability} states that realization \eqref{eq:maximum_state_space} has an unobservable subspace of initial conditions, i.e., there exist initial conditions whose effect cannot be observed in the output $y_k$ for any $k\in\Nee$. This unobservability results from the state definition \eqref{eq:maximum_ss_state_def}, in which inputs and outputs are separately included in the state. In contrast, in the SISO LTI case, minimal realizations such as the observable canonical form are constructed by defining state variables as a function of both inputs and outputs according to $x_{k-\ell} = y_{k-\ell} + u_{k-\ell}$, effectively combining $y_{k-\ell}$, $u_{k-\ell}$ with the same lag $\ell$ into the same state variable \cite[section 2.3.3]{Goodwin1984}, requiring only one state dimension. Thus, separately including inputs and outputs in the state is a source of nonminimality of realization \eqref{eq:maximum_state_space}. Note that the minimal state construction of the LTI case is not possible in the LPV case due to the time dependency of the scheduling signal \cite{Toth2007}. 

Next, Theorem \ref{th:observability} is specialized to the cases in which $n_b=1$ (inverse LPV-FIR) with realization \eqref{eq:maximum_state_space_invFIR} or $n_a=0$ (LPV-FIR) with realization \eqref{eq:maximum_state_space_FIR}.

\begin{corollary}
    State-space realization \eqref{eq:maximum_state_space_invFIR} is structurally $n_a$-observable if and only if there exists a $\bar{p}\in\Pee$ such that $\rank\ A_{n_a}(\bar{p}) = n_y$, and completely $n_a$-observable if and only if $\rank\ A_{n_a}(\bar{p}) = n_y$ for all $\bar{p}\in\Pee$. Last, \eqref{eq:maximum_state_space_invFIR} is always completely $n_a$-reconstructible.
    \end{corollary}
\begin{proof}
    These claims follow directly from the proof of Lemma \ref{lem:Ocalbar} and Theorem \ref{th:reconstructability} and \ref{th:observability} by substituting $\Fcal(p_j)$ and $\Hcal(p_j)$ by $(F_a - G_a A(p_j)$ and $-A(p_j)$. Specifically, $\bar{\Ocal}_k$ for \eqref{eq:maximum_state_space_invFIR} is given by
    \begin{equation}
        \bar{\Ocal}_k = \begin{bmatrix}
            -A(p_0) \\
            -A(p_1) F_a \\
            \vdots \\
            -A(p_{k-1}) F_a^{k-1}            
        \end{bmatrix} \in \Ree^{n_y k \times n_x},
    \end{equation}
    with $n_x = n_a n_y$. Then, for $k < n_a$, it holds that $\rank\ \bar{\Ocal}_k \geq n_y k < n_x $. If $k \geq n_a$, $\bar{\Ocal}_k$ can be written as
    \begin{equation}
        \bar{\Ocal}_k = \left[\begin{array}{ccccc}
            -A_1(p_0) & -A_2(p_0) & \hdots & -A_{n_a-1}(p_0) & -A_{n_a}(p_0) \\
            -A_2(p_1) & -A_3(p_1) & \hdots & -A_{n_a}(p_1) &  \\
            \vdots & & \raisebox{-3pt}{\rotatebox{70}{$\ddots$}} & & \\
            -A_{n_a-1}(p_{n_a-2} & -A_{n_a}(p_{n_a-2}) & & & \\
            -A_{n_a}(p_{n_a-1}) & & & & 0 \\
            \hline
            \multicolumn{5}{c}{0_{k - n_a \times n_x}}
        \end{array}\right].
    \end{equation}
    Consequently, by the block-triangular structure, $\rank\ \bar{\Ocal}_k \leq n_a n_y = n_x$ with equality holding if and only if $\rank\ A_{n_a}(p_j) = n_y$ for $j=0,\ldots,k-1$. Consequently,  \eqref{eq:maximum_state_space_invFIR} is completely observable if and only if $\rank\ A_{n_a}(\bar{p})) = n_y \ \forall \bar{p}\in\Pee$, and it is structurally observable if and only if there exists a $\bar{p} \in \Pee$ for which $\rank\ A_{n_a}(\bar{p})) = n_y$. Complete reconstructability follows by the same argument as Theorem \ref{th:reconstructability}.
\end{proof}
Above corollary links observability of \eqref{eq:maximum_state_space_invFIR} to the rank of the coefficient function $A_{n_a}$ related to the highest delayed output $y_{k-n_a}$. This can be understood by noting that if $A_{n_a}$ does not have full rank, some initial conditions for $y_{k-n_a}$ do not contribute to $y_k$ and are immediately shifted out of the state $\bar{y}_k$. Interestingly, this observability condition is the same as the reachability condition in Corollary \eqref{cor:reachability_invFIR}. Next, \eqref{eq:maximum_state_space_FIR} is considered.

\begin{corollary}\label{cor:FIR_observability}
    State-space realization \eqref{eq:maximum_state_space_FIR} with $n_y \geq n_u$ is structurally $(n_b-1)$-observable if there exists a $\bar{p}\in\Pee$ such that $\rank\ B_{n_b-1}(\bar{p}) = n_u$. It is completely $(n_b-1)$-observable if $\rank\ B_{n_b-1}(\bar{p}) = n_u\ \forall \bar{p}\in\Pee$. Last, \eqref{eq:maximum_state_space_FIR} is always completely $(n_b-1)$-reconstructible. State-space realization \eqref{eq:maximum_state_space_FIR} with $n_y < n_u$ is not observable.
\end{corollary}
\begin{proof}
    For realization \eqref{eq:maximum_state_space_FIR}, the observability matrix is directly given by
    \begin{equation}
        \Ocal_k = 
        \begin{bmatrix}
            B(p_0) \\
            B(p_1) F_b \\
            \vdots \\
            B(p_{k-1}) F_b^{k-1}
        \end{bmatrix} \in \Ree^{k n_y \times n_x},
    \end{equation}
    with $n_x =n_u (n_b-1)$. For $k = n_b-1$, $\Ocal_k$ is given by
    \begin{equation}\label{eq:cor_FIR_proof_Ok}
        \Ocal_{n_b-1} = 
        \left[\begin{array}{ccccc}
            B_1(p_0) & B_2(p_0) & \hdots & B_{n_b-2}(p_0) & B_{n_b-1}(p_0) \\
            B_2(p_1) & B_3(p_1) & \hdots & B_{n_b-1}(p_1) &  \\
            \vdots & & \raisebox{-3pt}{\rotatebox{70}{$\ddots$}} & & \\
            B_{n_b-2}(p_{n_b-3} & B{n_b-1}(p_{n_b-3}) & & & \\
            B_{n_b-1}(p_{n_b-2}) & & & & 0 \end{array}\right] \in \Ree^{n_y (n_b-1) \times n_u (n_b-1)},
    \end{equation}
    with $k > n_b-1$ only adding zero rows by the nilpotency of $F_b$. The anti block upper triangular structure of $\Ocal_{n_b-1}$ in \eqref{eq:cor_FIR_proof_Ok} immediately allows for imposing conditions on the rank of $B_{n_b-1}$ to guarantee full rank of $\Ocal_{n_b-1}$ as follows.
    \begin{enumerate}[leftmargin=*]
        \item If $n_y \geq n_u$, i.e., the coefficient functions $B_i$ are square or tall, also $\Ocal_{n_b-1}$ is tall. Then, if there exists a $\bar{p}\in\Pee$ such that $\rank\ B_{n_b-1}(\bar{p}) = n_u$, this implies that $\rank\ \Ocal_k = n_x$ for that $\bar{p}$, i.e., realization  \eqref{eq:maximum_state_space_FIR} is structurally $(n_b-1)$-observable. Moreover, if $\rank\ B_{n_b-1}(\bar{p}) = n_u \ \forall \bar{p}\in\Pee$, this implies that  $\rank\ \Ocal_k = n_x$ for all $p$, i.e., realization \eqref{eq:maximum_state_space_FIR} is completely $(n_b-1)$-observable.
        \item  If $n_y < n_u$, i.e., the coefficient functions $B_i$ are wide, also $\Ocal_{n_b-1}$ is wide, i.e., $n_x = n_u (n_b-1) > n_y (n_b-1)$. Then $\rank\ \Ocal_k < n_x$ for all $p$, and \eqref{eq:maximum_state_space_FIR} is not observable. \vspace{-20pt}%
    \end{enumerate} %
\end{proof}
When $n_y \geq n_u$, Corollary \ref{cor:FIR_observability} links observability of \eqref{eq:maximum_state_space_FIR} to the rank of the coefficient function $B_{n_b-1}$ related to the highest delayed input $u_{k-n_b-1}$. Intuitively, if $B_{n_b-1}$ does not have full rank, initial states associated to $u_{k-n_b+1}$, i.e., the last part of the state, cannot be observed. Note that these conditions are only sufficient, as for $n_y \geq n_u$, all initial conditions might be observed after $k< n_b-1$ time steps. Consider for example that $n_y = n_u (n_b-1)$ and $B_i(p_k)$ is a constant full rank matrix. Then all initial conditions are observed in the first output. Last, in case $n_y < n_u$, a wide $B_{n_b-1}$ does not allow for fully observing the initial states associated to $u_{k-n_b+1}$, such that in this setting \eqref{eq:maximum_state_space_FIR} is always unobservable.

To conclude, it is shown that the state-space realization \eqref{eq:maximum_state_space} resulting from the direct state construction \eqref{eq:maximum_ss_state_def} is not observable and thus not minimal for any coefficient functions $A_i,B_i$. However, it is completely $\max(n_a,n_b-1)$-reconstructible. In the (inverse) FIR setting, \eqref{eq:maximum_state_space_invFIR} and \eqref{eq:maximum_state_space_FIR} can be observable depending on the rank of the highest-order coefficient function $A_{n_a}$ or $B_{n_b-1}$. All findings are summarized in Tab. \ref{tab:summary}.

\begin{table}[t]
\caption{Summary of reachability, observability and reconstructability conditions.}
\label{tab:summary}
\resizebox{\textwidth}{!}{
\begin{tabular}{@{}lll@{}}
\toprule
$n_a>0,n_b>1$ & $n_a>0,n_b=1$ & $n_a=0,n_b>1$ \\ \midrule
\textbf{Structurally $n_x$-reachable} &  &  \\
\begin{tabular}[c]{@{}l@{}} $\textnormal{\rank} \begin{bmatrix} I + \sum_{i=1}^{n_a} \sigma^{-i}A_i(\bar{p}) \\ \sum_{i=0}^{n_b-1} \sigma^{-i} B_i(\bar{p}) \end{bmatrix}^\T = n_y \ \forall \sigma \in \Cee \backslash \{ 0 \}$ \\ $\textnormal{\rank} \begin{bmatrix} -A_{n_a}(\bar{p}) & B_{n_b-1}(\bar{p}) \end{bmatrix} = n_y$\end{tabular} & $\textnormal{\rank}\ B_0(\bar{p}) = n_y$ & Yes \\ \midrule
\textbf{Completely $k$-observable} &  &  \\
No & $k = n_a$ iff $\textnormal{\rank}\ A_{n_a}(\bar{p}) = n_y \ \forall \bar{p} \in \Pee$ & \begin{tabular}[c]{@{}l@{}}$n_y \geq n_u$: if $\textnormal{\rank}\ B_{n_b-1}(\bar{p}) = n_u \ \forall \bar{p} \in \Pee$, $k=n_b-1$ \\ $n_y < n_u$: no\end{tabular} \\ \midrule
\textbf{Structurally $k$-observable} &  &  \\
No & $k=n_a$ iff $\exists \bar{p} \in \Pee$ s.t. $\textnormal{\rank}\ A_{n_a}(\bar{p}) = n_y$ & \begin{tabular}[c]{@{}l@{}}$n_y \geq n_u$: if $\exists \bar{p} \in\Pee$ s.t. $\textnormal{\rank}\ B_{n_b-1}(\bar{p}) = n_u$, $k=n_b-1$ \\ $n_y < n_u$: no\end{tabular} \\ \midrule
\textbf{Completely $k$-reconstructible} &  &  \\
Yes, $k=\max(n_a,n_b-1)$ & Yes, $k=n_a$ & Yes, $k=n_b-1$ \\ \bottomrule
\end{tabular}
}
\end{table}

\begin{remark}
    By complete detectability of \eqref{eq:maximum_state_space}, the state $x_k$ of \eqref{eq:maximum_state_space} satisfies $\lim_{k\mapsto \infty} x_k = 0$ if and only if the output $y_k$ of \eqref{eq:LPV_IO_DT} satisfies $\lim_{k\mapsto \infty} y_k = 0$ since the only directions in which $x_k$ can be nonzero, decay to $0$ in $k=\max(n_a,n_b-1)$ steps by Theorem \ref{th:reconstructability}. Particularly, if considering the class of inputs $\{u \ | \ u_k = 0 \ \forall k > k_1 \} \cap \signalspaceDT$ for some $k_1 \in \Zee$, i.e., inputs that are cut off at $k_1$, complete detectability implies that $y_k$ satisfies $\lim_{k\mapsto \infty} y_k = 0$ if there exists a $\Pcal \in \See_{\succ 0}^{n_x}$ such that $\Fcal^\T(p_k) \Pcal \Fcal(p_k) - \Pcal \prec 0 \ \forall p_k\in\Pee$. If considering only a frozen point $\bar{p}\in\Pee$, i.e., in an LTI setting, this is also necessary, i.e., $\lim_{k\mapsto \infty} y_k = 0$ if and only if there exists a $\Pcal \in \See_{\succ 0}^{n_x}$ such that $\Fcal^\T(\bar{p}) \Pcal \Fcal(\bar{p}) - \Pcal \prec 0$.
\end{remark}

\section{Examples}
\label{sec:examples}
In this section, the main observability and reachability results in Theorems \ref{th:DT_controllability} and \ref{th:observability} are illustrated through academic examples\footnote{The code for these examples is available at \protect\url{https://gitlab.tue.nl/kon/nonminimum-lpv-ss-realizations/-/tree/main/DT}}. Specifically, the examples illustrate a non-exhaustive collection of mechanisms through which the realization \eqref{eq:maximum_state_space} can lose observability/reachability, and link these mechanisms to the conditions in Theorems \ref{th:DT_controllability} and \ref{th:observability}. The considered mechanisms are as follows.
\begin{enumerate}[noitemsep]
    \item Including previous inputs $u_{k-i}$ as separate states, as opposed to incorporating it in the state variables together with previous outputs, resulting in a loss of observability, illustrating Theorem \ref{th:observability}.
    \item Not taking into account shared directions in $A_i$, resulting in a loss of reachability, illustrating condition \eqref{eq:last_coefficient_nullspace} of Theorem \ref{th:DT_controllability}. In the LTI case, this corresponds to not taking into account the multiplicity of the pole.
    \item Incorporating states for coefficient functions $A,B$ that are not coprime in a functional sense, resulting in a loss of reachability, illustrating condition \eqref{eq:coprime_theorem}. In the LTI case, this corresponds to a pole-zero cancellation.
    \item A multi-input multi-output (MIMO) LTI example in which the true number of poles is not a multiple of the output dimension, resulting in pole-zero cancellations and a loss of reachability, illustrating condition \eqref{eq:coprime_theorem} of Theorem \ref{th:DT_controllability}.
\end{enumerate}

\subsection{Mechanism 1}\label{sec:mechanism1}
This example illustrates mechanism 1 and 3. Specifically, consider the IO representation
\begin{equation}\label{eq:example1_IO}
    y_k = -a_1(p_k) y_{k-1} - a_2(p_k) y_{k-2} + b_0(p_k) u_k + b_1(p_k) u_{k-1},
\end{equation}
with scheduling function $p: \Zee \mapsto \Pee = [1,\infty)$ and coefficient functions $a_1(p_k) = 2p_k$, $a_2(p_k) = p_k^2$, $b_0 = p_k$, and $b_1(p_k) = p_k^{-1}$.
The nonminimal state-space realization of \eqref{eq:example1_IO} is given by
\begin{subequations}\label{eq:example1_nonminSS}
\begin{align}
    \bar{x}_{k+1} &= \left[\begin{array}{c} y_k \\ y_{k-1} \\ \hline u_{k} \end{array}\right] = \left[\begin{array}{cc|c} -a_1(p_k) & -a_2(p_k) & b_1(p_k) \\ 1 & 0 & 0 \\ \hline 0 & 0 & 0 \end{array}\right] \left[\begin{array}{c}y_{k-1} \\ y_{k-2} \\ \hline u_{k-1}\end{array}\right] + \left[\begin{array}{c} b_0(p_k) \\ 0 \\ \hline 1 \end{array}\right] u_k \\
    y_k &= \left[\begin{array}{cc|c} -a_1(p_k) & -a_2(p_k) & b_1(p_k) \end{array}\right] \bar{x}_k + b_0(p_k) u_k,
\end{align}
\end{subequations}
with $\bar{x}_k \in \Ree^3$. Reachability and observability of this realization are discussed next.

\textbf{Reachability}: To conclude on reachability, Theorem \ref{th:DT_controllability} is applied. Specifically, first consider that $p_k = \bar{p} = 1 \ \forall k\in\Zee$, and consider the condition in \eqref{eq:coprime_theorem}. For $\bar{p}=1$, it holds that $1 + \sum_{i=1}^{n_a} a_i(\bar{p}) \sigma^{-i} = 1 + 2 \sigma^{-1} + 1 \sigma^{-2} = 0$ for $\sigma = -1$. Simultaneously, $\sum_{i=0}^{n_b-1} b_i(\bar{p}) \sigma^{-i} = 1 + 1 \sigma^{-1} = 0$ also holds for $\sigma=-1$, i.e., the polynomials associated with $a_i$ and $b_i$ have a common root $\sigma = -1$ for $\bar{p} = 1$. Consequently, $\begin{bmatrix} 1 + \sum_{i=1}^{n_a} \sigma^{-i}a_i(\bar{p}) & \sum_{i=0}^{n_b-1} \sigma^{-i} b_i(\bar{p}) \end{bmatrix} = 0$ for $\bar{p} = 1$ and $\sigma = -1$. Thus, \eqref{eq:example1_nonminSS} is not reachable at $\bar{p}=1$ by Theorem \ref{th:DT_controllability}. Second, consider that $p_k = \bar{p} = 2$. In this setting, $1 + \sum_{i=1}^{n_a} a_i(\bar{p}) \sigma^{-i} = 0$ holds for $\sigma = -2$. However, for $\sigma = -2$, $\sum_{i=0}^{n_b-1} b_i(\bar{p}) \sigma^{-i} = 2 + 0.5 (-2)^{-1} \neq 0$, such that \eqref{eq:example1_nonminSS} is reachable at $\bar{p}=2$ by Theorem \ref{th:DT_controllability}. Consequently \eqref{eq:example1_nonminSS} is structurally reachable.

\textbf{Observability}: To investigate observability, Theorem \ref{th:observability} is applied. Specifically, by Lemma \ref{lem:Ocalbar}, the $4$-step observability matrix $\Ocal_4 \in \Ree^{4 \times 3}$ of \eqref{eq:example1_nonminSS} is defined through
\begin{align}
    \left[\begin{array}{cccc} 1 & 0 & 0 & 0\\ a_{1}(p_{k+1}) & 1 & 0 & 0\\ a_{2}(p_{k+2}) & a_{1}(p_{k+2}) & 1 & 0\\ 0 & a_{2}(p_{k+3}) & a_{1}(p_{k+3}) & 1 \end{array}\right] \Ocal_4 
    =      
    \left[\begin{array}{ccc} -a_{1}(p_k) & -a_{2}(p_k) & b_{1}(p_k) \\ -a_{2}(p_{k+1}) & 0 & 0 \\ 0 & 0 & 0 \\ 0 & 0 & 0 \end{array}\right] = \bar{\Ocal}_4
 \nonumber
\end{align}
Inspecting $\bar{\Ocal}_4$ yields $\rank\ {\Ocal}_4= \rank\ \bar{\Ocal}_4 = 2 < n_x$ for any $p \in \signalspaceDT$ as $a_2(\bar{p}) \geq 1$ for any $\bar{p} \in \Pee = [1,\infty)$. By Lemma \ref{lem:Ocalbar}, $\rank\ \bar{\Ocal}_k \leq 2$ also for $k > 4$ as only zero rows are added to  $\bar{\Ocal}_k$. Thus, \eqref{eq:example1_nonminSS} is unobservable. However, it is reconstructible in 2 time steps by Theorem \ref{th:reconstructability}. 

To obtain a minimal realization for \eqref{eq:example1_IO}, a state-transformation is defined to project out the unobservable subspace. Specifically, define $v_3(p_k) = \begin{bmatrix} 0 & b_1(p_k) & a_2(p_k) \end{bmatrix}^\T$ as a basis for the unobservable subspace, i.e., $\ker\ \bar{\Ocal}_4 = \im\ v_3(p_k)$. Note that this basis depends on $p_k$. Furthermore, define $v_1 = \begin{bmatrix} 1 & 0 & 0 \end{bmatrix}^\T$ and $v_2(p_k) = \begin{bmatrix} 0 & -a_2(p_k) & b_1(p_k) \end{bmatrix}^\T$ as a basis for the observable subspace, i.e., $(\ker\ \bar{\Ocal}_4)^\perp = \im\begin{bmatrix} v_1(p_k) & v_2(p_k) \end{bmatrix}$. Then define $T(p_k) = \begin{bmatrix} v_1 & v_2(p_k) & v_3(p_k) \end{bmatrix}$, for which it is claimed that $\rank\ T(p_k) = 3$ for any $p_k\in\Pee$. To see this, note that $v_1,v_2,v_3$ are mutually orthogonal, i.e., $v_i^\T (p_k) v_j(p_k) = 0 \ \forall p_k \in\Pee$ for $i=1,2,3$, $j=1,2,3$. A state transformation $x_k = T(p_k) z_k$ and thus $x_{k+1} = T(p_{k+1}) z_{k+1}$ yields the equivalent state-space representation
\begin{subequations} \label{eq:example1_minSS}
\begin{align}
    z_{k+1} &= \left[\begin{array}{cc|c}
        -a_1(p_k) & \gamma(p_k) & 0 \\
        -a_2(p_{k+1}) \gamma^{-1}(p_{k+1}) & 0 & 0 \\
        \hline b_1(p_{k+1}) \gamma^{-1}(p_{k+1}) & 0 & 0
    \end{array} \right] z_k + 
    \begin{bmatrix}
        b_0(p_k) \\ b_1(p_{k+1}) \gamma^{-1}(p_{k+1}) \\ \hline a_2(p_{k+1}) \gamma^{-1}(p_{k+1})
    \end{bmatrix} u_k \\
    y_k &=
    \left[\begin{array}{cc|c} -a_1(p_k) & \gamma(p_k) & 0 \end{array} \right] z_k + b_0(p_k) u_k
\end{align}
\end{subequations}
in which $\gamma(q) = a_2(q)^2 + b_1(q)^2$. In this transformed representation, the last state $z_{k,3}$ is not observable. Consequently, a minimal realization with state dimension $2$ for \eqref{eq:example1_IO} is obtained by deleting the last rows/columns of \eqref{eq:example1_minSS}.

To conclude, this example illustrates that including the delayed input $u_{k-1}$ in a separate state results in an unobservable subspace, and that this can be avoided by, for example, combining $u_{k-1}$ and $y_{k-2}$ into a single state as prescribed by $v_2$. 

\begin{remark}
    Note that even though \eqref{eq:example1_nonminSS} is not minimal, its coefficient functions only depend on $p_k$, whereas the coefficient functions of \eqref{eq:example1_minSS} also depend on shifts of the scheduling signal $p_{k+1}$, referred to as a dynamic dependency. Furthermore, the coefficient functions of \eqref{eq:example1_nonminSS} depend affinely on $a_i,b_i$, whereas the coefficient functions of \eqref{eq:example1_minSS} are a nonlinear functions of the original $a_i,b_i$. This nonlinear, dynamic dependency on $a_i,b_i$ complicates stability/dissipativity analysis and controller synthesis, such that in these settings, it can be beneficial to employ the nonminimal realization \eqref{eq:example1_nonminSS}.
\end{remark}

\subsection{Mechanism 2}
This example illustrates mechanism 2 in an LTI setting. Specifically, consider the LTI IO representation
\begin{equation}\label{eq:example2_IO}
    y_k = -\begin{bmatrix} .3 & 0 \\ 0 & .3 \end{bmatrix} y_{k-1}  + \begin{bmatrix} 3 \\ 1 \end{bmatrix} u_k = -A_1 y_{k-1} + B_0 u_k,
\end{equation}
for which the corresponding transfer function is given by
\begin{equation}\label{eq:example2_TF}
    y_k = \begin{bmatrix}
        \frac{3}{1 + .3 z^{-1}} \\
        \frac{1}{1 + .3 z^{-1}} 
    \end{bmatrix}.
\end{equation}
A minimal realization for \eqref{eq:example2_TF} is obtained as
\begin{align}\label{eq:example2_minSS}
    \hat{x}_{k+1} = -0.3 \hat{x}_{k} + u_k & & y_k = \begin{bmatrix} -0.9 & -0.3 \end{bmatrix}^\T x_k + \begin{bmatrix} 3 & 1 \end{bmatrix}^\T.
\end{align}
Since both entries of \eqref{eq:example2_TF} contain the same pole at $z = -0.3$, this minimal realization has $\hat{x}_k \in \Ree$, i.e., only one state is required for this pole. In contrast, the nonminimal realization for \eqref{eq:example2_IO} is given by
\begin{align}\label{eq:example2_nonminSS}
    \bar{x}_{k+1} = -A_1 \bar{x}_k + B_0 u_k & & y_k = -A_1 x_k + B_0 u_k
\end{align}
with $\bar{x}_k \in \Ree^2$, which has two uncoupled states that could have been combined into a single state.

The nonminimality of \eqref{eq:example2_nonminSS} manifests through an unreachable subspace of $\bar{x}_k \in \Ree^2$. Specifically, by Corollary \ref{cor:reachability_invFIR}, realization \eqref{eq:example2_minSS} is reachable if $\rank\ B_0 = n_y = 2$. However, $\rank\ B_0 = 1$, such that \eqref{eq:example2_nonminSS} is not reachable. Indeed, the reachability matrix $\Rcal_k = \begin{bmatrix} B_0 & 0.3 B_0 & 0.3^2 B_0 & \hdots & 0.3^{k-1} B_0 \end{bmatrix}$ satisfies $\rank\ \Rcal_k = 1$ for any $k$. The unreachable subspace is thus given by $(\im\ B_0)^\perp = \textrm{span} \begin{bmatrix} 1 & -3 \end{bmatrix}^\T$. Last, note that in the LTI case, Corollary \ref{cor:reachability_invFIR} is not only sufficient for reachability, but also necessary \cite{hespanha2018linear}.

\subsection{Mechanism 3}
This example illustrates mechanism 3. Specifically, consider the IO representation
\begin{equation}\label{eq:example3_IO}
    y_k + p_{k} y_{k-1} = u_k + p_k u_{k-1},
\end{equation}
with $p_k\in\Pee = \Ree$. A nonminimal state-space realization is given by
\begin{align}\label{eq:example3_nonminSS}
    \left[\begin{array}{c}
        y_k \\ \hline u_k
    \end{array}\right]
    =
    \left[\begin{array}{c|c}
        -p_{k} & p_k \\ \hline 0 & 0
    \end{array}\right]
    \left[\begin{array}{c}
        y_{k-1} \\ \hline u_{k-1}
    \end{array}\right]
    + \left[\begin{array}{c}
        1 \\ \hline 1
    \end{array}\right] u_k 
    & & y_k = \left[\begin{array}{c|c} -p_{k} & p_k \end{array}\right] x_k + u_k
\end{align}
with $x_k\in\Ree^2$.

\textbf{Reachability:} By Theorem \ref{th:DT_controllability}, \eqref{eq:example3_nonminSS} is reachable if there exists a $\bar{p}\in\Pee$ such that $\rank \begin{bmatrix} 1 + \bar{p}\sigma^{-1} & 1 + \bar{p}\sigma^{-1}\end{bmatrix} = 1 \ \forall \sigma \in \Cee$ and $\rank \begin{bmatrix} -\bar{p} & \bar{p} \end{bmatrix} = n_y$. However, such a $\bar{p}$ does not exist as the first condition fails if $\bar{p} \neq 0$ for $\sigma = -\bar{p}$, and the second one if $\bar{p} = 0$. Consequently, reachability of \eqref{eq:example3_nonminSS} cannot be concluded from Theorem \ref{th:DT_controllability}. This can be understood by noting that for a constant scheduling $p_k = \bar{p}\in\Pee \ \forall k\in\Zee$, both sides of \eqref{eq:example3_IO} associate with the same polynomial $1+\bar{p} \sigma^{-1}$ and are thus not coprime. In other words, \eqref{eq:example3_IO} corresponds to transfer function $G(z) = (1 + \bar{p} z^{-1})(1 + \bar{p} z^{-1})^{-1}$ for which a pole-zero cancellation occurs for all frozen LTI behaviors. 

Since Theorem \ref{th:DT_controllability} is only sufficient for concluding reachability, also the  reachability matrix is analyzed, which for $k=3$ is given by
\begin{equation}
    \Rcal_3 = \begin{bmatrix} \Fcal(p_{k+2}) \Fcal(p_{k+1}) \Gcal & \Fcal(p_{k+2}) \Gcal & 
    \Gcal \end{bmatrix} = \begin{bmatrix} -p_{k+2} (-p_{k+1} + p_{k+1}) & -p_{k+2} + p_{k+2} & 1 \\ 0 & 0 & 1 \end{bmatrix} = \begin{bmatrix} 0 & 0 & 1 \\ 0 & 0 & 1 \end{bmatrix},
\end{equation}
such that $\rank\ \Rcal_3 = 1$. More generally, the reachable subspace is given by $\im\ \Rcal_k = \im\ \begin{bmatrix} 1 & 1\end{bmatrix}^\T$ for any $p\in\signalspaceDT$ since $\Fcal(p_{k+i}) \Gcal = 0$ for any $p_{k+i}$. Thus \eqref{eq:example3_nonminSS} is unreachable and its unreachable subspace is $p$-independent and given by $\im\ \begin{bmatrix} 1& -1\end{bmatrix}^\T$. This can be intuitively understood by noting that from rest, i.e., for $y_{k-1}=u_{k-1} =0$, it holds by \eqref{eq:example3_IO} that $y_k = u_k$. Consequently, in the next time steps, $p_k y_{k-1}$ and $p_k u_{k-1}$ cancel and again $y_k = u_k$, thus resulting in the given unreachable subspace. 

\textbf{Observability:} The observability matrix of \eqref{eq:example3_nonminSS} is defined by 
\begin{equation}
    \begin{bmatrix} 1 & 0 & 0 \\ p_{k+1} & 1 & 0 \\ 0 & p_{k+2} & 1 \end{bmatrix}\Ocal_3 = \begin{bmatrix} -p_k & p_k \\ 0 & 0 \\ 0 & 0 \end{bmatrix},
\end{equation}
for $k=3$ such that $\rank \ \Ocal_3 = 1 < n_x$. Generally, by Lemma \ref{lem:Ocalbar}, for any $k$ it holds that $\rank \ \Ocal_k = 1$ with the unobservable subspace given by $\ker\ \Ocal_k = \im\ \begin{bmatrix} 1 & 1 \end{bmatrix}^\T$.


\subsection{Mechanism 4}
This example illustrates Theorem \ref{th:DT_controllability} in the MIMO LTI setting and is adapted from \cite{Alsalti2023}. It is shown that if the true amount of poles is not a multiple of $n_y$, then an IO model can contain extra poles which have to be canceled by zeros, resulting in non-coprimeness of $I+\sum_{i=0}^{n_a} A_i \sigma^{-i}$ and $\sum_{i=0}^{n_b-1} B_i \sigma^{-i}$, violating condition \eqref{eq:coprime_theorem} of Theorem \ref{th:DT_controllability}, such that the nonminimal realization \eqref{eq:maximum_state_space} of the IO model is not reachable.

Specifically, consider a MIMO LTI IO representation with
\begin{align}\label{eq:example4_IO_1}
    y_k \!=\! -\underbrace{\left[\begin{array}{cc} 0.435 & -1.52\\ 0.802 & 0.074 \end{array}\right]}_{A_1} y_{k-1} - \underbrace{\left[\begin{array}{cc} -0.584 & -0.272\\ 1.938 & 1.524 \end{array}\right]}_{A_2} y_{k-2} + \underbrace{\left[\begin{array}{cc} 0.1 & -0.3\\ -0.1 & -0.7 \end{array}\right]}_{B_1}u_{k-1} \underbrace{\left[\begin{array}{cc} 0.286 & -0.294\\ -1.097 & 1.267 \end{array}\right]}_{B_2} u_{k-2}.
\end{align}
A direct nonminimal realization of \eqref{eq:example4_IO_1} can directly be constructed using $A_1,A_2,B_1,B_2$ as in \eqref{eq:maximum_state_space} with $n_a=2,n_b=3$, resulting in an eight-dimensional state.

To test reachability of the direct nonminimal realization, Theorem \ref{th:DT_controllability} is applied, stating that this realization is reachable if
\begin{align}\label{eq:example4_reachability}
    \rank \begin{bmatrix} I + \sum_{i=1}^{n_a} A_i \sigma^{-i} & \sum_{i=0}^{n_b} B_i \sigma^{-i} \end{bmatrix} = n_y \ \forall \sigma \in \Cee \backslash \{ 0 \} & & \textrm{and} & &
    \rank \begin{bmatrix}
        -A_2 & B_2
    \end{bmatrix} = n_y.
\end{align}
The second condition is verified as $\rank\ A_2 = n_y$. To address the first condition, note that $\rank( P(\sigma)) = \rank (I + \sum_{i=1}^{n_a} A_i \sigma^{-i}) < n_y$ only for $\sigma \in \mathfrak{S} = \{ 0.3406 \pm 1.7314i, -1.2806, 0.09066\}$, while $\rank\ P(\sigma) = n_y \ \forall \sigma \backslash \mathfrak{S} $, hence the first condition only has to be tested at $\sigma \in \mathfrak{S}$. This yields that $\rank \begin{bmatrix} I + \sum_{i=1}^{n_a} A_i \sigma^{-i} & \sum_{i=0}^{n_b} B_i \sigma^{-i} \end{bmatrix} = 1 < n_y$ for $\sigma = 0.09066$. Since Theorem 3 is also necessary in the LTI case, the direct nonminimal realization is not reachable and has a one-dimensional unreachable subspace. Intuitively, this can be understood by noting that $Q(\sigma) = \sum_{i=0}^{n_b} B_i \sigma^{-i}$ also has a zero at $\sigma = 0.09066$.

To test observability of the realization, Lemma \ref{lem:Ocalbar} is applied. By this lemma, it holds that the observability matrix $\Ocal_8$ of this direct realization satisfies $\rank\ \Ocal_8 = 4$, resulting in a four-dimensional unobservable subspace.

A minimal realization for the IO representation \eqref{eq:example4_IO_1} is given by \cite{Alsalti2023}
\begin{align}\label{eq:SS_example4}
    x_{k+1} = \left[\begin{array}{ccc} -0.5 & 1.4 & 0.4\\ -0.9 & 0.3 & -1.5\\ 1.1 & 1 & -0.4 \end{array}\right] x_k + \left[\begin{array}{cc} 0.1 & -0.3\\ -0.1 & -0.7\\ 0.7 & -1 \end{array}\right] u_k & & 
    y_k =  \left[\begin{array}{ccc} 1 & 0 & 0\\ 0 & 1 & 0 \end{array}\right] x_k,
\end{align}
i.e., with a state dimension of $n_x=3$, matching with the four-dimensional unobservable and one-dimensional unreachable subspaces of the direct nonminimal realization with a state dimension of eight. 

To conclude, in this example it is shown that if the true amount of poles ($3$) is not a multiple of the output dimension ($2$),  $P(\sigma)$ almost always includes an extra pole that has to be canceled by $Q(\sigma)$ \cite{Gevers1986}, resulting in non-coprime $P(\sigma)$ and $Q(\sigma)$. Consequently, the direct nonminimal realization is not reachable. Thus, in the MIMO setting, it is important to have dedicated parametrizations for IO models that match the true order of the system.

\section{Conclusion}
In this paper, a direct state-space realization for LPV input-output representations is provided and numerical conditions are developed to test its observability and reachability properties. It is shown that this direct realization is nonminimal. However, it is always reconstructible and under coprimeness and well-posedness conditions, this realization is also reachable. 

As a consequence, this state-space realization can immediately be used to verify stability/dissipativity properties of an LPV-IO model obtained using system identification. Additionally, it can be used in several LMI-based controller and observer synthesis methods and recently developed LPV data-driven controller design methods, as the unobservable modes are stable by themselves.

\bibliographystyle{IEEEtran}
\bibliography{IEEEabrv,BSTcontrol.bib,library.bib}

\end{document}